\documentclass[pre,twocolumn,aps,superscriptaddress, floatfix]{revtex4-2}
\usepackage[T1]{fontenc}
\usepackage[utf8]{inputenc}

\usepackage{array}
\usepackage{mathrsfs}
\usepackage{multirow}
\usepackage{amsmath}
\usepackage{amsthm}
\usepackage{amssymb}
\usepackage{mathbbol}
\usepackage{graphicx}
\usepackage{latexsym}
\usepackage{textcomp}
\usepackage{mathtools}
\usepackage{xcolor}
\usepackage{braket}
\usepackage{physics}

\usepackage[citecolor=magenta,colorlinks=true]{hyperref}
\usepackage{bm}
\definecolor{mycolor1}{rgb}{0.1, 0.6, 0.6}
\usepackage{calligra} \DeclareMathAlphabet{\mathcalligra}{T1}{calligra}{m}{n} \DeclareFontShape{T1}{calligra}{m}{n}{<->s*[2.2]callig15}{}

\renewcommand{\ij}{\langle ij \rangle}
\newcommand{\ijplus}{\langle ij \rangle_{\bm{\hat{e}}}}

\newcommand{\bhat}[1]{\bm{\hat{#1}}}
\renewcommand{\d}{\textrm{d}}

\begin{document}

\title{Green's Functions for Random Resistor Networks}

\author{Sayak Bhattacharjee}
\email{sayakb@iitk.ac.in}
\affiliation{%
Department of Physics, Indian Institute of Technology Kanpur, Kanpur 208016, India}%
\author{Kabir Ramola}%
\email{kramola@tifrh.res.in}
\affiliation{%
Tata Institute of Fundamental Research, Hyderabad 500107, India}%

\date{\today}

\begin{abstract}
We analyze random resistor networks through a study of lattice Green's functions in arbitrary dimensions. We develop a systematic disorder perturbation expansion to describe the weak disorder regime of such a system. We use this formulation to compute ensemble averaged nodal voltages and bond currents in a hierarchical fashion. We verify the validity of this expansion with direct numerical simulations of a square lattice with resistances at each bond exponentially distributed. Additionally, we construct a formalism to  recursively obtain the {\it exact} Green's functions for finitely many disordered bonds. We provide explicit expressions for lattices with up to four disordered bonds, which can be used to predict nodal voltage distributions for arbitrarily large disorder strengths. Finally, we introduce a novel order parameter that measures the overlap between the bond current and the optimal path (the path of least resistance), for a given resistance configuration, which helps to characterize the weak and strong disorder regimes of the system. 
\end{abstract}

\maketitle

\section{Introduction} \label{section_introduction}
Electrical networks have often been used to model a wide variety of condensed matter phenomena, 
both in steady-state as well as transient regimes. While they have recently arisen as synthetic experimental test-beds for topological quantum matter~\cite{bandstructure_prb, topoelectric_circuits, topological_prl}, a longstanding application of electrical networks---in particular, resistor networks---has been to model the conductivity of disordered random media. This problem of fundamental interest is relevant for transport measurements \cite{rrn_graphene, percolationtransition_afrydman, Altshuler_PRL,  Basko_PRB} and also the study of critical phenomena \cite{cardy_transition, stephen_meanfield, Conglio_RRN_PRB, Lubensky_noise_PRB}.  In order to model natural systems, microscopic disorder in such model systems is an important ingredient. Such situations often require a subtle understanding of the properties of the lattice Green's function (which is related to the inverse of the lattice Laplacian) \cite{Bradley_PRE, Bhagavatula_PRE}, and therefore an investigation into techniques that can be used to compute Green's functions for disordered electrical networks represents a fundamental direction of theoretical as well as experimental relevance. 

Within this context, random resistor networks (RRN) are a popular paradigm for modeling transport in disordered media such as semiconductors \cite{skirkpatrick_rrn, rrn_graphene, percolation_randommedia, z_ball_rrn, ambegaonkar_halperin, kirkpatrick_prl, bernasconi, doussal, percolation_model_strelnicker}, but also in directed polymers \cite{banavar_optimalpaths, banavar_invasionpercolation, ahansen_optimalpaths}, and porous rocks \cite{barabasi_porousrocks, stanley_postbreakthrough_behaviour}. Formally defined, an RRN is a network of resistors such that the resistances are sampled from a probability distribution, and the disorder strength may be controlled by a tunable parameter \cite{abharris_rrn_percolation, halperin_RRN, halpinhealy}. Studies of such networks can be approached through a multitude of ways, most primarily by percolation theory  \cite{kirkpatrick_prl, berman_percolation, skirkpatrick_rrn, frydman_percolationmodel, percolationtransition_afrydman, abharris_rrn_percolation, angulo_conductance_distributions, Coniglio, shklovskiui1975percolation}, but also through random walks \cite{tetali_randomwalks, jeng_randomwalks_rrn}, and optimization theory \cite{mst_on_randomnetworks, optimalpaths_in_disorderedmedia, banavar_optimalpaths, stanley_optimalpaths}. 

While transport in directed polymers and porous rocks is classical, transport in mesoscopic systems can often be in the quantum (ballistic) regime, where the conductances are described by the Landauer-B\"uttinker formalism \cite{datta1997electronic}. However, we investigate RRNs in the Ohmic limit, which is relevant for mesoscopic transport in the \textit{diffusive regime}, whenever the conduction length exceeds the mean free path and phase coherence length of the electronic wavefunction \cite{datta1997electronic}, as achieved in various contexts \cite{ramakrishnan2017equivalence, caprara2011effective, suarez2020two}. In fact, this formalism is valid whenever the DC resistance can be defined independent of the voltage between the terminals, which is also achieved in quantum transport whenever $eV$ and $k_\textrm{B}T$ are smaller than the transmission coefficient \cite{bagwell1989landauer} (see for instance Refs. \cite{dash2022comprehensive, bao2011modeling, kim2023continuous}).

In this work, we consider an RRN with resistances sampled from an exponentially wide distribution \cite{halperin_RRN, stanley_pre, bernasconi}. This model is also termed as the \textit{hopping percolation model} \cite{frydman_percolationmodel, percolationtransition_afrydman, percolation_model_strelnicker} and can be motivated from physical considerations: the conductance between sites in disordered media is often proportional to $\textrm{exp}\left(-r_{ij}/r_0-E_{ij}/k_\textrm{B}T\right)$ where $r_0$ is a length scale for the decay of the wavefunction of the grains, $r_{ij}$ is the distance and $E_{ij}$ is the energy difference between two sites $i$ and $j$. Thus, an exponential disorder of the form $\textrm{exp}(ax_{ij})$ is natural: the strength $a$ represents an energy and/or length scale of the disordered system corresponding to the random variable $x_{ij}$ \cite{percolation_model_strelnicker}. 

In exponentially disordered networks, past studies have identified two disorder regimes---a \textit{weak} disorder regime when $L\gg a^\nu$ and a \textit{strong} disorder regime for $L\ll a^\nu$, where $\nu$ is the percolation connectedness exponent ($\nu=4/3$ in two dimensions) \cite{stauffer_aharony, stanley_pre, percolation_model_strelnicker, frydman_percolationmodel, percolationtransition_afrydman}. The strong disorder regime is characterized by optimal behavior: the current distribution collapses to a self-similar fractal \textit{optimal path} \cite{stanley_optimalpaths, stanley_pre} (see Fig.~\ref{fig:current_dist} for a visualization), whose critical exponents ($d_\textrm{opt}=1.22$ in two dimensions) have been numerically computed \cite{banavar_optimalpaths}. In the weak disorder regime, the current distribution is delocalized throughout the lattice and the optimal path (defined as the path of least resistance) is shown to be self-affine with critical exponents belonging to the universality class of directed polymers \cite{stanley_optimalpaths, optimalpaths_in_disorderedmedia}. This crossover from self-affine to self-similar is a characteristic of wide exponential disorder; studies for Gaussian and uniform distributions give self-affine optimal paths across disorder strength \cite{optimalpaths_selfaffine}. Optimal paths have, of course, been understood as equivalent to domain walls in spin systems as well, where the impurities of the system help pin the wall to energetically favourable sites in the system \cite{huse_henley,kardar, huse_henley_fisher}. While critical exponents of the disorder regimes have been explored in depth, a scalable analytical toolbox to analyse such disordered networks has not yet been developed.

In this article, we fill this gap by demonstrating the use of two analytic techniques to compute the lattice Green's functions of the disordered system. First, we construct a perturbation theory that provides a hierarchical expansion for the nodal voltages in powers of the disorder variables. Perturbative expansions have been attempted in prior literature for computing the lattice conductance \cite{skirkpatrick_rrn, jmluck_conductivity}, or in lattices with a small number of disordered bonds \cite{cserti_perturbation}, however, our formulated expansion helps to explicitly compute disorder-averaged system observables such as nodal voltages and bond currents by solving the Dyson equation order by order. Similar perturbation expansions for Green's functions have been helpful in studying disordered crystalline media \cite{acharya2020athermal,das2021long,pappu_disorderperturbationexpansion, pappu_emergentpowerlaw, debankur_disorderednetworks, roshan_athermalfluctuationsthree, roshan_firstcontact}.

Next, we develop an exact formulation using a dyadic perturbation of the lattice Green's function, that can in principle yield exact results for arbitrary disorder. The dyadic bond formulation enables us to provide exact formulae for the Green's function for a finite number of disordered bonds in the lattice. Although similar ideas were adopted previously for a single broken bond in the system \cite{cserti_perturbation}, we extend such a formulation to an arbitrary number of bonds {\it with} disorder and are able to give an analytically tractable expression for lattices with a small number of disordered bonds. In particular, we provide explicit formulae for lattices with up to four disordered bonds. 

We also perform numerical simulations that corroborate our theoretical results, as well as help us probe the different disorder regimes of the system. We compute disorder-averaged nodal voltages with one, two and three disordered bonds in the system, and study their fluctuations. The fluctuations are shown to peak at a critical disorder strength and our numerics match perfectly with the analytical predictions from exact formulae for the Green's functions. We also provide a novel order parameter, which we term \textit{bond current fidelity}. This is defined as the overlap between the current distribution at a particular disorder strength and the optimal path for a particular resistance configuration. We demonstrate clear signatures of the weak and strong disorder regimes in this order parameter, and its scalings are shown to be in line with previously known critical exponents and our analytical predictions using the Green's function formalism. 

\begin{figure}[t!]
    \includegraphics[width=0.6\columnwidth]{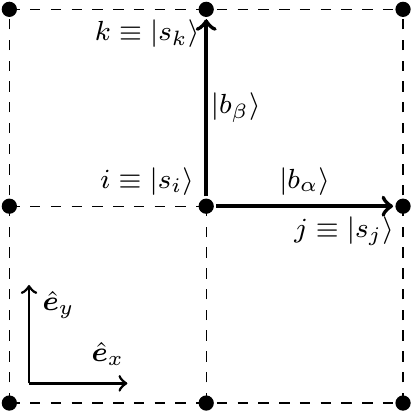}
    \caption{The lattice convention used in this paper. Each lattice site $i$ is represented by a $N_s=L^d$ dimensional column vector $| s_i \rangle$, that forms an orthonormal basis. The basic lattice translation vectors are denoted by $\{\bm{\hat{e}}_m\}$ with $(1\leq m\leq d)$ and to each site $i$, we assign the $d$ bonds along $\bm{\hat{e}}$, as shown using the solid arrows along the bonds. Two representative bond vectors are thus, $\ket{b_\alpha}\equiv\ket{s_i}-\ket{s_j}$ and $\ket{b_\beta}\equiv \ket{s_i}-\ket{s_k}$, where $1\leq \alpha,\beta\leq N_b=2L^d$.
    }
    \label{fig:lattice_convention}
\end{figure}

The rest of the paper is organized as follows. In Sec.~\ref{sec:rrn} we introduce the random resistor model and its lattice Laplacian formulation. We discuss the disorder perturbation expansion for the Green's function and compute the nodal voltages perturbatively in Sec.~\ref{sec:perturbationexpansion}. In Sec.~\ref{sec:dyadicbond}, we introduce the exact formulation to obtain Green's functions for arbitrarily many bonds with disorder in the lattice. This work is supplemented by numerical techniques for nodal voltages and discussion of an order parameter in Sec.~\ref{sec:numerics}. Finally, we discuss and conclude the work in Sec.~\ref{sec:discussion}, and present directions for future research.

\section{Random resistor network}\label{sec:rrn}

In this work, we consider a $d$-dimensional hypercubic lattice of linear dimension $L$ with periodic boundary conditions, and resistors placed at each bond. There are $L^d \:\: (=N_s)$ sites and $d L^d \:\: (=N_b)$ bonds on such a $d$-dimensional torus. While we consider a hypercubic lattice with periodic boundary conditions, our formulation can be easily adapted to other lattices with alternate boundary conditions as well. Each lattice site is denoted by a site index (in Latin alphabets) $i\equiv (x_1,x_2,\hdots,x_d)$ where $\{x_k\}$ $(1\leq k\leq d)$ are the Cartesian coordinates of the real space vector corresponding to site $i$ ($1\leq i\leq N_s$).
For the formulation developed in this paper, we find it convenient to use bra-ket notation to denote the total degrees of freedom on the lattice. The bras and kets are vectors in $N_s$ dimensional space. We define a basis set of site vectors $\{ \ket{s_i} \}$ that denote the sites so that the vector $\ket{s_i}$ (denoted equivalently by site index $i$) is the $i^{\textrm{th}}$ unit vector in $N_s$ dimensional space. Thus, $\bra{s_j}\ket{s_i}=\delta_{ij} \:\: (1\leq i,j\leq N_s)$ and the site vectors form a complete orthonormal set, that is, $\sum_{i}\ket{s_i}\bra{s_i}=1$. A site index subscript on a ket denotes the corresponding entry of the column vector, for example, $\ket{s_i}_{j}$ is the $j^{\textrm{th}}$ entry of this column vector. The notation $\ij$ indicates that sites $i$ and $j$ are connected by a bond on the lattice (nearest neighbours). Analogous to the site vectors, it is also useful to define bond vectors. We denote a (positive) unit vector along a Cartesian axis by $\bhat{e}$. Then, to each site $i$, we unambiguously associate the $d$ bonds along the positive Cartesian axes and denote the oriented bond along $\bhat{e}$ by the notation $\ijplus$. We can now define the bond vectors for the bonds in the lattice (indexed by Greek letters) by
\begin{equation}
\ket{b_\alpha}\coloneqq\ket{s_i}-\ket{s_j} \;\;\textrm{with}\;\; \ijplus.
\end{equation}
The bond index $\alpha$ ($1\leq \alpha\leq N_b$) is equivalent to the tuple $(i,\bm{\hat{e}})$ 
which uniquely marks the bond starting at $i$ along $\bhat{e}$ (see Fig. \ref{fig:lattice_convention}).

In the network, each of the bonds between two sites $i$ and $j$ has a resistance of magnitude $R_{ij}$. For the voltages at each site, we define a nodal voltage vector $\ket{V}$. We also define a nodal current vector $\ket{I}$, which represents the algebraic sum of currents exiting a site. At all sites $i$ in the lattice not connected to external leads, Kirchhoff's current law trivially demands $\ket{I}_i=0$. Thus, typically, the nodal voltages are the most relevant (unknown) degrees of freedom. To denote the bond currents, we define $d$ bond current vectors $\{\ket{J_{\bm{\hat{e}}}}\}$ corresponding to the current flowing in the $\bm{\hat{e}}$ direction in the relevant bond assigned to each site. The $N_b$-dimensional complete bond current vector is denoted by $\ket{J}\equiv \ket{J_{\bm{\hat{e}}_1}|J_{\bm{\hat{e}}_2}|\hdots |J_{\bm{\hat{e}}_d}}$, where $|$ separates the $d$ blocks of the ket.

Our formulation in the following sections is independent of the specificities of the external current or voltage configuration that sets up a steady-state in the lattice (which we solve for). In conductivity measurements, a popular choice is a bus-bar configuration, where the voltages are fixed on two opposite sides \cite{rrn_graphene}. In our numerics, we consider a simpler configuration---we fix a current source node $i_\textrm{in}$ and current sink node $i_{\textrm{out}}$ which input and output unit current respectively. We thus set the nodal current configuration to be given by
\begin{equation}
    \ket{I}_{i}\coloneqq \delta_{i,i_{\textrm{in}}}-\delta_{i,i_{\textrm{out}}},\label{cur_def},
\end{equation}
without loss of generality. We find this choice particularly useful to demonstrate current localization to optimal paths in the circuit.

\begin{figure}[t!]
    \includegraphics[width=\columnwidth]{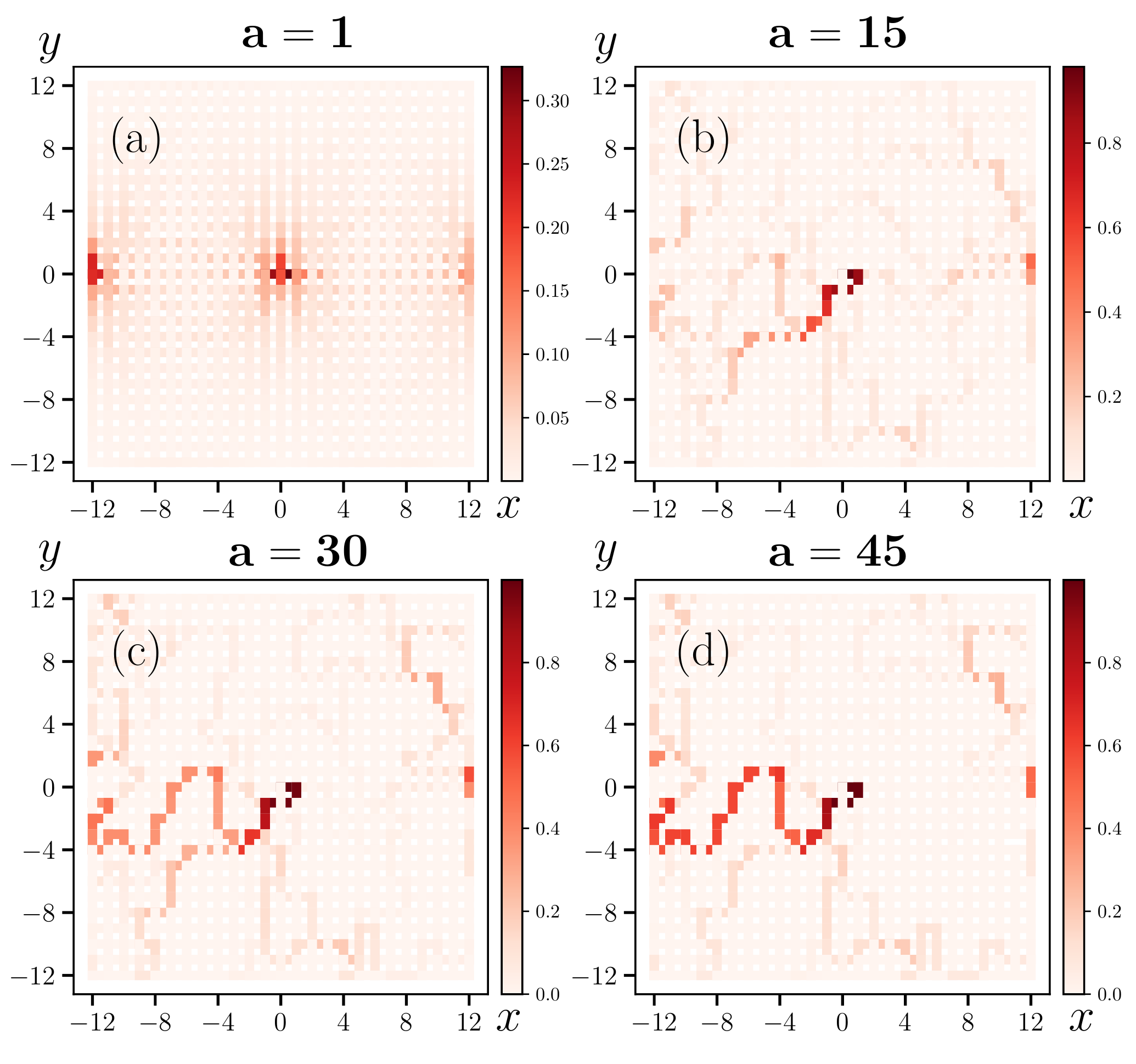}
    \caption{Current distributions for a $25 \times 25$ lattice with exponential disorder (see Sec.~\ref{sec:perturbationexpansion}) at disorder strengths of (a) $a=1$, (b) $a=15$, (c) $a=30$ and (d) $a=45$. The resistances at each bond are distributed as  $R_{ij} \coloneqq e^{ax_{ij}}$, with the random variables $\{x_{ij}\}$ remaining fixed as the disorder strength $a$ is increased. The current source and sink are at $(-12,0)$ and $(0,0)$ respectively. Convergence to an optimal path (dark red path in (d)) can be clearly observed with increasing disorder strength.}
    \label{fig:current_dist}
\end{figure}

As stated before, the theory holds for any arbitrary voltage or current source-sink configuration. In this particular work, we present simulations for $i_{\textrm{in}}\equiv (-(L-1)/2,0)$ and $i_{\textrm{out}}\equiv (0,0)$ on the square lattice in two dimensions, so that the current enters at the midpoint of the left boundary and exits at the center node of the lattice (see Fig.~\ref{fig:current_dist} for reference). This source-sink configuration forces the system size $L$ to be an odd integer.

\subsection*{Lattice Laplacian Formulation}
We assume that the bond resistances in the RRN are independent of the voltage difference between the sites (as in Ohm's law), and obtain the following,
\begin{equation}
\ket{J_{\bm{\hat{e}}}}_{i}=\frac{\ket{V}_{i}-\ket{V}_{j}}{R_{ij}} \:\:\textrm{with}\:\:\ijplus .\label{ohmlaw}
\end{equation}
Now, due to local charge conservation in the steady state, we apply Kirchhoff's current law, given by
\begin{equation}
\ket{I}_i=\sum_{j \:\textrm{with}\:\langle ij\rangle}\frac{\ket{V}_i-\ket{V}_j}{R_{ij}}.  \label{kcl}
\end{equation}
where $j$ with $\ij$ indicates a sum over index $j$ whenever $i$ and $j$ are connected by a bond.
We study RRNs where the bond resistances are perturbed from a mean resistance $R_0$, which without loss of generality we can set equal to 1 unit. We can recast Eq.~\eqref{kcl} in the following linear algebraic form
\begin{equation}
    \textsf{L}\ket{V}+\ket{I}=0, \label{lvi}
\end{equation}
where $\textsf{L}$ is the lattice Laplacian (or conductance matrix) \cite{cserti_greensfunction_ajp}. Explicitly, the Laplacian is given by
\begin{equation}
      [\textsf{L}]_{ij}\coloneqq \begin{cases}-\sum_{j\:\textrm{with}\:\ij} (R_{ij})^{-1} &  \textrm{if}\; i=j\\
      (R_{ij})^{-1} &   \textrm{if}\; \ij \\
0 & \textrm{otherwise}.
\end{cases}\label{laplacian_def}
\end{equation}
Observe that when all resistances are equal to $R_0$, $\textsf{L}$ reduces to the usual circulant form of the lattice Laplacian of a $d$-dimensional torus, as expected. Clearly, Eq.~\eqref{lvi} can be solved by inverting the Laplacian, thus $\ket{V}=-\textsf{L}^{-1}\ket{I}$. Thus our basic object of study is the lattice Green's function given by
\begin{equation}\label{greens_intro}
\textsf{G} \equiv -\textsf{L}^{-1},
\end{equation}
which provides all the system properties. We must be careful to note that due to the sum rule implemented by Kirchoff's current law \cite{percolation_elastic}, the Laplacian is a non-invertible matrix, and hence, must be inverted by projecting out the zero mode $\ket{0}=(1\:1\:\hdots 1)^T$ of the Laplacian. Specifically, the Green's function and the Laplacian are related by
\begin{equation}    \textsf{L}\textsf{G}=\textsf{G}\textsf{L}\equiv -(\mathbb{1}-\ket{0}\bra{0}).\label{gl_correpondence}
\end{equation}
Since the voltages in the system are equivalent upto an arbitrary constant, this Green's function can be used as an inverse without concern. While, in principle, such an inversion may be performed numerically, in Sec. \ref{sec:perturbationexpansion} and \ref{sec:dyadicbond}, we construct analytic techniques to compute the Green's function for a disordered lattice in terms of the Green's function for the perfect lattice. 

We now demonstrate how to compute the bond currents generally. The relationship in Eq.~\eqref{ohmlaw} may be recast into the following linear algebraic form
\begin{equation}  \textsf{D}_{\bm{\hat{e}}}\ket{V}+\ket{J_{\bm{\hat{e}}}}=0,\label{dvj}
\end{equation}
where the difference matrices are given explicitly by 
\begin{equation}
[\textsf{D}_{\bm{\hat{e}}}]_{ij}\coloneqq \left(R_{\ijplus}\right)^{-1}\begin{cases}-1 & \textrm{if} \;\; i=j \\
    1 &\textrm{if} \;\; \ijplus \\  0&\textrm{otherwise}\end{cases}\label{difference_def}
\end{equation}
where $R_{\ijplus}$ is the resistance on $\ijplus$. Again, notice that when all the resistances are equal to $R_0$, this generalized difference matrix becomes the usual difference operator for a $d$-dimensional torus. Thus, once the voltages are known, the bond currents can be simply computed using $\ket{J_{\bhat{e}}}=-\textsf{D}_{\bhat{e}}\ket{V}$. 

Our formulations in the following sections will be independent of the explicit choice of disorder in the resistances. As motivated in the introduction, however, we intend to study the crossover from the regimes of weak to strong disorder in the hopping percolation model~\cite{stanley_pre}, which is obtained by setting
\begin{equation}
    R_{ij}\coloneqq e^{ax_{ij}},\label{res_exp_dis}
\end{equation}
where $x_{ij} \in  (0,1)$ (and $i$ and $j$ share a bond) is a uniformly distributed random variable and $a$ controls the strength of the disorder. The limit $a\rightarrow 0$ yields a lattice with zero disorder (perfect lattice), while $a\rightarrow \infty$ provides the strong disorder limit. For ease of analytical calculations, we find it convenient to introduce the scalar variables $\{\zeta_{ij}\}$ which represent the disorder in the bond resistances. The explicit relationship considered is given by
\begin{equation}
    R_{ij}\coloneqq (1-\zeta_{ij})^{-1}.\label{res_def}
\end{equation}
Then, from Eqs.~\eqref{res_exp_dis} and ~\eqref{res_def}, one can find that the distribution of the variables $\zeta$ is given by (for $a\geq 0$)
\begin{equation}
    f(\zeta)= a^{-1}(1-\zeta)^{-1} \;\;\; \textrm{for} \;\; 0<\zeta <1- e^{-a}.
\end{equation}
This also implies that the resistances obey an inverse probability distribution, that is, $f(R)=1/(aR)$ for $1\leq R\leq e^a$. The moments of the disorder $\zeta$ can also be computed exactly; in particular, the first three moments are given by $\langle \zeta \rangle =(-1+a+e^{-a})/a$, $\langle \zeta^2 \rangle =(-3+2a+4e^{-a}-e^{-2a})/(2a)$, and $\langle \zeta^3 \rangle =(-11+6a+18e^{-a}-9e^{-2a}+2e^{-3a})/(6a)$ where $\langle \cdot\rangle$ denotes a disorder ensemble average.

\section{Disorder Perturbation Expansion}\label{sec:perturbationexpansion}

\begin{figure*}[t!]
    \includegraphics[width=\textwidth]{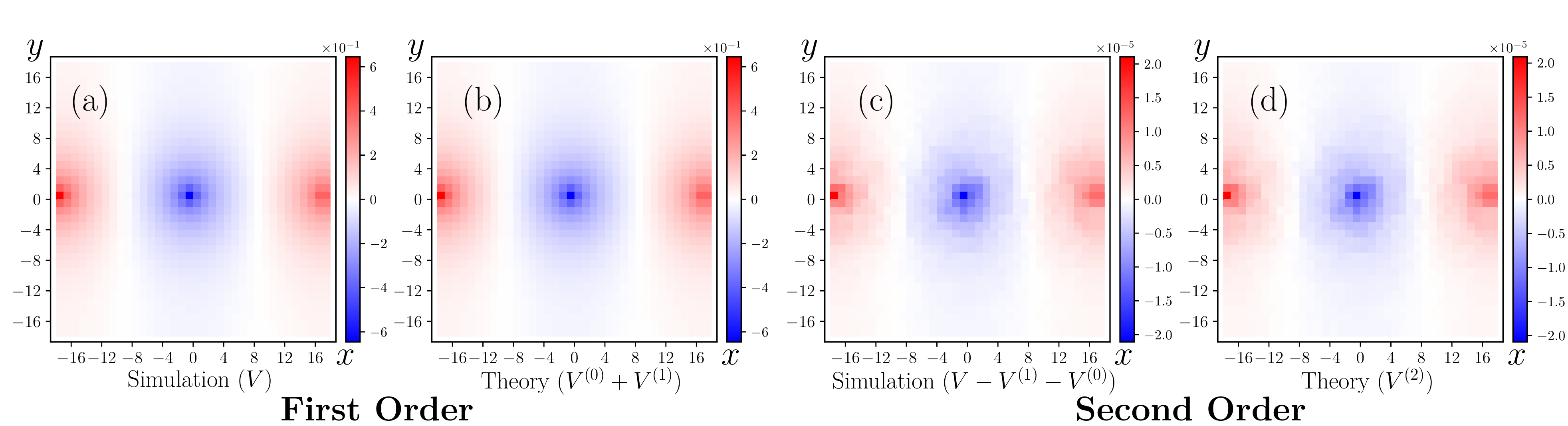}
    \caption{Nodal voltages in a $35 \times 35$ lattice with periodic boundary conditions. The current source and sink are at $(-17,0)$ and $(0,0)$ respectively. The disorder strength is $a=0.01$. The plots depict (a) nodal voltages obtained from the simulations $\ket{V}$, (b) nodal voltages predicted from the theory upto linear order $\ket{V}^{(0)}+\ket{V}^{(1)}$, (c) nodal voltages obtained from the difference of the simulations and the theoretical predictions upto first order $\ket{V}-[\ket{V}^{(0)}+\ket{V}^{(1)}]$, and (d) nodal voltages predicted by the theory at second order $\ket{V}^{(2)}$. The numerical results and theoretical predictions match exactly at first order ((a) and (b)) as well as second order ((c) and (d)). Note the difference in the magnitudes between the solutions at first and second order.}
    \label{fig:voltage_diff_0.01}
\end{figure*}
For the weak disorder regime, that is, a small deviation from the perfect lattice, we can compute the degrees of freedom accurately using a perturbation expansion in the disorder. We control such a perturbation by a tuning parameter $\lambda$, such that $0<\lambda<1$ and $\lambda=0$ corresponds to the zero disorder state. Therefore, we associate the tuning parameter to the disorder $\zeta$ so that the resistances are redefined as $ R_{ij}\equiv (1-\lambda\zeta_{ij})^{-1}$. We consider linear perturbations on the Laplacian and difference operators as follows
\begin{subequations}
 \begin{align}
\textsf{L}&\coloneqq \textsf{L}^{(0)}+\lambda \textsf{L}^{(1)}, \label{perturbation in laplacian} \\
\textsf{D}_{\bhat{e}}&\coloneqq \textsf{D}_{\bhat{e}}^{(0)}+\lambda \textsf{D}_{\bhat{e}}^{(1)}. 
\end{align}\label{perturbation in difference matrix}
\end{subequations}
Note, the explicit forms of the perfect lattice's Laplacian $\textsf{L}^{(0)}$ and difference operator $\textsf{D}_{\bhat{e}}^{(0)}$ can be obtained from Eqs.~\eqref{laplacian_def} and \eqref{difference_def} by setting all resistance magnitudes $R_{ij}$ to 1.

A linear order perturbation in these operators of the network is complete, and should induce perturbation expansions (in $\lambda$) upto all higher orders for the system variables. Thus, we assume
\begin{subequations}
\begin{align}
\ket{V}&\coloneqq \ket{V}^{(0)}+\lambda \ket{V}^{(1)} +\lambda^2 \ket{V}^{(2)}+\mathcal{O}(\lambda^3), \label{voltage expansion}\\
\ket{J_{\bhat{e}}}&\coloneqq \ket{J_{\bhat{e}}}^{(0)}+\lambda \ket{J_{\bhat{e}}}^{(1)} +\lambda^2  \ket{J_{\bhat{e}}}^{(2)}+\mathcal{O}(\lambda^3)\label{current expansion}
    \end{align}
\end{subequations}
where the superscript denotes the order of the expansion; naturally, the $(0)$ index denotes the values of the quantities in the zero disorder state. 

Applying the constraint that the above equations must obey Ohm's and Kirchhoff's laws at each order of $\lambda$, we obtain a hierarchical scheme to explicitly determine the higher order corrections to each of the above quantities. We first use Eqs.~\eqref{kcl} and \eqref{lvi} to determine the corrections to the Laplacian matrices and the nodal voltages. Using Eq.~\eqref{res_def} in Eq.~\eqref{kcl}, we have the following exact relationship between the currents and the nodal voltages, given a realization of the disorder
\begin{equation}
\ket{I}_{i}=\sum_{j\:\textrm{with}\:\ij}\left(\ket{V}_{i}-\ket{V}_{j}\right)\left(1-\lambda\zeta_{ij}\right).
\end{equation}
Next, comparing this with Eq.~\eqref{lvi}, we notice that
\begin{equation}
     [\textsf{L}^ {(1)}]_{ij}=\begin{cases}\sum_{j\:\textrm{with}\:\ij}\zeta_{ij}& \textrm{if} \; i=j\\
-\zeta_{ij} &\textrm{if}\;  \ij\\
0 &\textrm{otherwise}.\end{cases} \label{perturbed_laplacian}
\end{equation}
Now putting Eqs.~\eqref{voltage expansion} and \eqref{voltage expansion} in Eq.~\eqref{lvi}, and equating the terms at each order of $\lambda$, we obtain the explicit form of the voltage expansion in Eq.~\eqref{voltage expansion}. The contribution of each order is obtained in a hierarchical fashion using the contribution of the previous order, and we obtain
\begin{equation}
    \ket{V} =\textsf{G}^{(0)}[\mathbb{1} +\lambda \textsf{L}^{(1)}\textsf{G}^{(0)}+\lambda^2(\textsf{L}^{(1)} \textsf{G}^{(0)})^2 +\mathcal{O}(\lambda^3)]\ket{I}, \label{volt_explicit_green}
\end{equation}
where $\mathbb{1}$ denotes the identity matrix. Here,  $\textsf{G}^{(0)}$ is the perfect lattice Green's function of the lattice Laplacian $\textsf{L}^{(0)}$. Explicitly, the value of the perfect lattice Green's function is given by the $d$-dimensional integral 
\begin{equation}
   [\textsf{G}^{(0)}]_{ij}\coloneqq\int_{-\pi}^\pi \frac{\textrm{d}x_1}{2\pi} \hdots \int_{-\pi}^\pi \frac{\textrm{d}x_d}{2\pi}\frac{e^{i(l_1x_1+l_2x_2+\hdots)}}{2\sum_{i=1}^d(1-\cos x_i)},
\end{equation}
where $|i-j|\coloneqq (l_1, l_2,\hdots,l_d)$. This integral can generally be evaluated numerically, however, there also exist closed-form expressions and recurrence formulae for $d=2$ and $3$, which are summarized in Ref.~\cite{cserti_greensfunction_ajp}. In this study, however, we find it simpler to evaluate the perfect Green's function numerically by using Eq.~\eqref{gl_correpondence}.

The complete disordered Green's function $\textsf{G}$ satisfies the Dyson equation, given by
\begin{equation}
    \textsf{G}=\textsf{G}^{(0)}+\textsf{G}^{(0)}\textsf{L}^{(1)}\textsf{G}.
\end{equation}
Thus, the disordered Laplacian $\textsf{L}^{(1)}$ is the so-called self-energy of the system.
It is important to note that as the perturbation expansion in Eq.~\eqref{perturbation in laplacian} terminates at order $1$, the above equation, is in principle exact at all orders. However, inverting such an equation is in general hard, and therefore is usually solved in a perturbative manner, as we proceed to do.
Note that if we set $\lambda=1$, and write $\ket{V}=\textsf{G}\ket{I}$, where $\textsf{G}$ is the Green's function for the disordered system, then we recover the Dyson series given by
\begin{equation}
    \textsf{G} =\textsf{G}^{(0)}+ \textsf{G}^{(0)}\textsf{L}^{(1)}\textsf{G}^{(0)}+\textsf{G}^{(0)}\textsf{L}^{(1)} \textsf{G}^{(0)}\textsf{L}^{(1)} \textsf{G}^{(0)} + \hdots.  \label{dyson_series}
\end{equation}
The $n^{\textrm{th}}$ term of the Dyson series gives us the $n^{\textrm{th}}$ order correction to the Green's function, as is evident from the form we obtained in Eq.~\eqref{volt_explicit_green}. Computing the nodal voltages is thus simply a task of computing the terms of the Dyson series order by order, which is trivial by the explicit knowledge of the disordered Laplacian as given in Eq.~\eqref{perturbed_laplacian}. 

One can also compute the corrections to the difference matrices and the bond currents. Observe that Eq.~\eqref{ohmlaw} can be rewritten as follows
\begin{equation}     \ket{J_{\bhat{e}}}_{i}=(\ket{V}_{i}-\ket{V}_{j})(1-\lambda\zeta_{ij}) \:\: \textrm{with}\:\:\ijplus,
\end{equation}
and comparing this with Eq.~\eqref{dvj}, we obtain
\begin{equation}
[\textsf{D}^{(1)}_{\bhat{e}}]_{ij}=\zeta_{\ijplus}\begin{cases}1 & \textrm{if} \;\; i=j \\
-1 &\textrm{if} \;\; \ijplus\\
0&\textrm{otherwise}.\end{cases}\label{perturbed_difference}
\end{equation}
Now putting Eqs.~\eqref{perturbation in difference matrix} and \eqref{current expansion} in Eq.~\eqref{dvj}, and again computing in a hierarchical fashion, we obtain the explicit form of the bond current expansion in Eq.~\eqref{current expansion} as follows
\begin{align}
   \nonumber \ket{J_{\bhat{e}}}=&-\Big[\textsf{D}_{\bhat{e}}^{(0)}\textsf{G}^{(0)}+\lambda\Big(\textsf{D}_{\bhat{e}}^{(0)}\textsf{G}^{(0)}\textsf{L}^{(1)}\textsf{G}^{(0)}+\\ 
   \nonumber &\textsf{D}_{\bhat{e}}^{(1)}\textsf{G}^{(0)}\Big)+\lambda^2\Big(\textsf{D}_{\bhat{e}}^{(0)}\textsf{G}^{(0)}\textsf{L}^{(1)} \textsf{G}^{(0)}\textsf{L}^{(1)} \textsf{G}^{(0)}+\\
    &\textsf{D}_{\bhat{e}}^{(1)} \textsf{G}^{(0)}\textsf{L}^{(1)}\textsf{G}^{(0)}\Big)+\mathcal{O}(\lambda^3)\Big]\ket{I}. \label{cur_explicit}
\end{align}

The perturbation expansion developed above should lead to an exact answer for the voltage (resp. current) at a given site (resp. bond) for any \textit{small} value of the disorder. The small value of the disorder is controlled by the scale after which the perturbation expansion is divergent, and this disorder scale is estimated in Sec.~\ref{connection with perturbation theory}. In convergent cases, one only needs to consider the first few terms in the expansion, which rapidly decay at large orders. We illustrate the exact convergence of such an expansion scheme in a $35\times 35$ lattice with exponentially distributed disorder in Fig. \ref{fig:voltage_diff_0.01}. In Fig.~\ref{fig:voltage_diff_0.01}(a) and (b), we demonstrate the exact matching between the perturbation expansion up to first order with the numerically obtained (simulation) nodal voltages, and in Fig.~\ref{fig:voltage_diff_0.01}(c) and (d), we show that the difference between the above two is exactly reproduced by the second order contribution. We find that an excellent match is found between the theoretical and numerical results as evident from the nodal voltage distributions.

We mention that a similar disorder perturbation expansion was attempted by Kirkpatrick~\cite{skirkpatrick_rrn} and by Derrida and Luck~\cite{derrida1983diffusion}. We have provided a detailed exposition and confirmed the validity of such a perturbation scheme through an explicit demonstration of the various terms in the expansion for both nodal voltages and bond currents. Below, we demonstrate the applicability of such techniques for RRNs through the computation of disorder averages. This involves introducing a dyadic formulation to compute moments of the nodal voltages and bond currents. In the next Section, we also introduce a recursive scheme that can be used to, in principle, compute the {\it exact} Green's function for such a system.

\subsection*{Ensemble averages}\label{sec:ensembleaverage}

The perturbation expansion formalism developed above provides the nodal voltages and bond currents for a particular resistance configuration in the lattice. From experimental considerations, it is more helpful to understand the behaviour of the system in an averaged fashion over disorder; we thus compute the disorder ensemble averages of the nodal voltages and bond currents as a function of the moments of the disorder $\zeta$. The averaging is performed over the disorder ensemble with each instance of the ensemble representing a resistance configuration generated from an independent sampling of the disorder distribution. We first compute the ensemble averages of the nodal voltages. Observing the Dyson series (Eq.~\eqref{dyson_series}), we realize that this problem reduces to calculating the ensemble averages of matrix products of alternating disordered Laplacians and perfect lattice Green's function, that is, quantities of the kind $\langle  \textsf{L}^{(1)}\textsf{G}^{(0)}\textsf{L}^{(1)}\hdots\textsf{G}^{(0)}\textsf{L}^{(1)}\rangle$, with $2r-1$ matrices for the $r^{\textrm{th}}$ order contribution. Clearly, from inspection of the disordered Laplacian (see Eq.~\eqref{perturbed_laplacian}), we notice that $\langle \textsf{L}^{(1)}\rangle=-\langle \zeta\rangle \textsf{L}^{(0)}$. To compute the higher order contributions efficiently, we utilize a convenient dyadic bond representation of the perturbed Laplacian. Thus, notice that the perfect and perturbed lattice Laplacians can be recast in the following form

\begin{subequations}
\begin{align}
\label{clean Laplacian dyadic}\textsf{L}^{(0)}&=-\sum_{\alpha=1}^{N_b}\ket{b_\alpha}\bra{b_\alpha},\\
\textsf{L}^{(1)}&=\sum_{\alpha=1}^{N_b}\zeta_{\alpha}\ket{b_\alpha}\bra{b_\alpha}. \label{laplacian_correction dyadic}
\end{align}
\end{subequations}
Notice the generality of the formalism in Eq.~\eqref{laplacian_correction dyadic}. By attaching a correction $\zeta_\alpha$ to any bond $\ket{b_\alpha}$ in the lattice, one can construct the lattice Laplacian of any arbitrary (disordered or ordered in some fashion) resistor network using the simple formula $\textsf{L}=\textsf{L}^{(0)}+\textsf{L}^{(1)}$. Its further usefulness will be expanded upon in Sec.~\ref{sec:dyadicbond}. 

To compute the aforementioned matrix product averages, we will require moments of the disorder variables on the bonds, which implies that they obey a cluster averaging scheme. Thus, we recall that $\langle \zeta_\alpha\zeta_{\beta}\rangle\equiv \langle \zeta^2\rangle \delta_{\alpha\beta}+\langle \zeta\rangle ^2(1-\delta_{\alpha\beta})$ and $\langle \zeta_\alpha\zeta_{\beta}\zeta_{\gamma}\rangle=\langle \zeta^3\rangle\delta_{\alpha\beta}\delta_{\alpha\gamma}+3\langle \zeta^2\rangle\langle \zeta\rangle\delta_{\alpha\beta}(1-\delta_{\alpha\gamma})+\langle \zeta\rangle ^3(1-\delta_{\alpha\beta})(1-\delta_{\alpha\gamma})$. Therefore, we obtain the following averages
\begin{subequations}
\label{correlations}
\begin{align}
&\label{one point correlation}\langle \textsf{L}^{(1)}\rangle=-\langle \zeta \rangle \textsf{L}^{(0)},\\ 
 \nonumber \label{two-point-correlation}
&\langle \textsf{L}^{(1)}\textsf{G}^{(0)}\textsf{L}^{(1)}\rangle=\kappa_2(\zeta)\sum_{\alpha}\mathcal{G}^0_{\alpha\alpha}\ket{b_\alpha}\bra{b_\alpha}+\\
&\hspace{2cm}\langle \zeta\rangle^2\sum_{\alpha,\beta}\mathcal{G}^0_{\alpha\beta}\ket{b_\alpha}\bra{b_{\beta}},
\\
 \label{three-point-correlation}
\nonumber 
&\langle \textsf{L}^{(1)}\textsf{G}^{(0)}\textsf{L}^{(1)}\textsf{G}^{(0)}\textsf{L}^{(1)}\rangle=\kappa_3(\zeta)\sum_{\alpha}(\mathcal{G}^0_{\alpha\alpha})^2\ket{b_\alpha}\bra{b_\alpha}+\\
\nonumber & \hspace{1.5cm}(3\langle \zeta^2\rangle\langle \zeta\rangle-2\langle \zeta\rangle^3) \sum_{\alpha,\beta}\mathcal{G}^0_{\alpha\alpha}\mathcal{G}^0_{\alpha\beta}\ket{b_\alpha}\bra{b_{\beta}}+\\
 &\hspace{1.5cm}\langle \zeta\rangle^3\sum_{\alpha,\beta,\gamma}\mathcal{G}^0_{\alpha\beta}\mathcal{G}^0_{\beta\gamma}\ket{b_\alpha}\bra{b_{\beta}},
\end{align}
\end{subequations}
where we define a dressed Green's function between two bonds given by $\mathcal{G}^0_{\alpha\beta}\equiv\mel{b_\alpha}{\textsf{G}^{(0)}}{b_{\beta}}$. The cumulants of the random variable $\zeta$ are $\kappa_2(\zeta)\equiv \langle \zeta^2\rangle -\langle \zeta \rangle^2$ and $\kappa_3(\zeta)\equiv \langle \zeta^3\rangle -3\langle \zeta^2 \rangle\langle \zeta \rangle+2\langle \zeta\rangle ^3$. Notice, the first order correction (Eq.~\eqref{one point correlation}) reproduces the result we predicted simply by inspection of the matrices as well. 

The ensemble averages for the bond currents can be computed in a similar fashion. The perfect and the perturbed lattice difference matrix can also be written in a dyadic representation as follows
\begin{subequations}
\begin{align}
 \textsf{D}^{(0)}_{\bhat{e}}&=-\sum_{\alpha=1}^{N_b}\ket{s_{i(\alpha)}}\bra{b_\alpha},\\
\textsf{D}^{(1)}_{\bhat{e}}&=\sum_{\alpha=1}^{N_b}\zeta_{\alpha}\ket{s_{i(\alpha)}}\bra{b_\alpha}.
\end{align}
\end{subequations}
where $\ket{s_{i(\alpha)}}$ is the site vector associated with the site $i$ as well as the bond $\ket{b_\alpha}$ such that $\alpha\equiv (i, \bhat{e})$. As evident from Eq.~\eqref{cur_explicit}, the relevant quantities to calculate in this case are again matrix products of alternating perfect lattice Green's function and perturbed Laplacian, except that the first matrix is the disordered difference matrix, that is, quantities of the kind  $\langle  \textsf{D}^{(1)}_{\bhat{e}}\textsf{G}^{(0)}\textsf{L}^{(1)}\hdots\textsf{G}^{(0)}\textsf{L}^{(1)}\rangle$. Owing to the similar structure of the dyadic representation, it is clear that the disorder averaged quantities are exactly the ones computed in Eqs.~\eqref{correlations}, with any dyad $\ket{b_\alpha}\bra{b_{\beta}}$ replaced by $\ket{s_{i(\alpha)}}\bra{b_{\beta}}$ on the right hand site and the initial $\textsf{L}^{(1)}$ replaced by $\textsf{D}^{(1)}_{\bhat{e}}$ in the matrix products, and $\textsf{L}^{(0)}$ replaced by $\textsf{D}^{(0)}_{\bhat{e}}$ on the left hand side of the equation. 

Given the computed averaged quantities in Eqs.~\eqref{correlations}, it is straightforward to compute the disorder averaged cumulants of the nodal voltages or bond currents. 
For example, using the Dyson series for the voltages and series expansion for the current in Eq.~\eqref{volt_explicit_green} and ~\eqref{cur_explicit} we obtain explicit series expansions for the first moment up to second order
\begin{align}
\nonumber \langle \ket{V}\rangle  =&[\textsf{G}^{(0)}+\lambda \textsf{G}^{(0)}\langle \textsf{L}^{(1)}\rangle \textsf{G}^{(0)}\\
&+\lambda^2\textsf{G}^{(0)}\langle \textsf{L}^{(1)} \textsf{G}^{(0)}\textsf{L}^{(1)} \rangle \textsf{G}^{(0)} +\mathcal{O}(\lambda^3)]\ket{I},\\
 \nonumber \langle \ket{J_{\bhat{e}}}\rangle =&-\Big[\textsf{D}_{\bhat{e}}^{(0)}\textsf{G}^{(0)}+\lambda\Big(\textsf{D}_{\bhat{e}}^{(0)}\textsf{G}^{(0)}\langle \textsf{L}^{(1)}\rangle \textsf{G}^{(0)}+\\ 
   \nonumber &\langle \textsf{D}_{\bhat{e}}^{(1)}\rangle \textsf{G}^{(0)}\Big)+\lambda^2\Big(\textsf{D}_{\bhat{e}}^{(0)}\textsf{G}^{(0)}\langle \textsf{L}^{(1)} \textsf{G}^{(0)}\textsf{L}^{(1)} \rangle \textsf{G}^{(0)}+\\
    &\langle \textsf{D}_{\bhat{e}}^{(1)} \textsf{G}^{(0)} \textsf{L}^{(1)}\rangle\textsf{G}^{(0)}\Big)+\mathcal{O}(\lambda^3)\Big]\ket{I}, \end{align}
and the relevant averages have been provided in Eq.~\eqref{correlations}. It is also useful to notice that these disorder averages are general for any arbitrary disorder distribution. 
Higher ($r^{\textrm{th}}$) order corrections are also straightforward to calculate---they would involve $r$-sums and the cluster averaged $r^{\textrm{th}}$ moments of the disorder variables. A convenient way to compute these correlation functions systematically may be via a diagrammatic expansion and can be formulated, in principle, for this setup as well. A typical route to diagrammatics is through a cumulant expansion in Fourier space~\cite{yonezawa_cpa, gonis_cpa, cpa_odagaki}, which can also be attempted within our formulation, however, we do not pursue such computations in this study (a similar attempt at diagrammatics has been pursued in Ref.~\cite{jmluck_conductivity}).

\section{Recursive dyadic bond disorder}\label{sec:dyadicbond}

While the perturbative expansion gives us accurate results upto the desired order in the weak disorder regime, it is unable to explain the behaviour of the system at strong disorder values. This is because the higher-order corrections are of comparable magnitudes, and therefore cannot be ignored. This is a caveat in using a perturbative approach, and hence we now present an alternative formulation for computing disordered Green's functions for this problem exactly. Although we demonstrate the explicit applicability of this technique for a small number of bonds with disorder in the lattice, instead of the scenario with disorder in all bonds studied before, the technique is general and can be extended to the aforementioned system as well. 

We denote the Green's function of the system when there are $n$ bond impurities in the lattice by $\textsf{G}^{[n]}$ (thus, $\textsf{G}^{[0]}$ is the perfect lattice Green's function, equivalent to $\textsf{G}^{(0)}$ in the perturbative framework). Note the distinction in the notation with the perturbative corrections to the Green's function---$\textsf{G}^{[n]}$ (square brackets in superscript) denotes the lattice Green's function for the system with $n$ disordered bonds, while $\textsf{G}^{(n)}$ (round brackets in superscript) denotes the $n^{\textrm{th}}$ order correction to the lattice Green's function (with disorder in all bonds) in the disorder perturbation expansion. The mathematical trick we employ is to use the Sherman-Morrison formula for matrices given by
\begin{equation}
    (\textsf{A}+\ket{u}\bra{v})^{-1}=\textsf{A}^{-1}-\frac{\textsf{A}^{-1}\ket{u}\bra{v}\textsf{A}^{-1}}{1+\mel{v}{\textsf{A}^{-1}}{u}}.
    \label{eq_sherman_morrison}
\end{equation}
where $\textsf{A}$ is an invertible square matrix and $u$ and $v$ are column vectors of appropriate size~\cite{shermanmorrison}. Such ideas were demonstrated by Cserti \textit{et al.}~\cite{cserti_perturbation} for the case of a single broken bond in the lattice. In this section we extend such a procedure to incorporate disorder in multiple bonds of the lattice.
As illustrated before in Sec.~\ref{sec:ensembleaverage}, each bond $\alpha$ can be represented in a dyadic form in terms of its bond vectors $\ket{b_\alpha}$. Earlier, our bond vectors were free of disorder, however, in this formulation, it is convenient to work with bond vectors with the disorder variable attached; thus, we define
\begin{equation}
\ket{\tilde{b}_{\alpha}}\equiv \sqrt{\zeta_\alpha}\ket{b_\alpha},
\end{equation}
where $\zeta_\alpha$ is the disorder variable on the $\alpha^{\textrm{th}}$ disordered bond. Note, here the index on the bond vector is not its index on the overall lattice, but instead locates it among the $n$ disordered bonds, with $1 \leq \alpha \leq n\leq N_b$. Now, the above matrix identity may be used iteratively (with $\textsf{A}$ replaced by the lattice Laplacian $\textsf{L}$) to obtain the disordered Green's functions for finitely many bond impurities $n$.

The central idea is a recursive application of the Sherman Morrison formula in Eq.~\eqref{eq_sherman_morrison} to substitute perfect bonds by disordered bonds step-by-step to construct the full disordered lattice. Consider the addition of a single disordered bond in the lattice, which modifies the Laplacian of a perfect lattice as follows
\begin{equation}
\textsf{L}^{[1]} = \textsf{L}^{[0]} + \ket{\tilde{b}_{1}}\bra{\tilde{b}_{1}},
\label{one_bond_laplacian}
\end{equation}
which should also be evident from Eq.~\eqref{laplacian_correction dyadic}. Here, in line with the notation for the Green's functions, a square bracket subscript on the lattice Laplacian denotes the network with finitely many disordered bonds. Similarly, adding a second disordered bond to the lattice modifies the Laplacian matrix of a system with an additional disordered bond, so that $\textsf{L}^{[2]} = \textsf{L}^{[1]} + \ket{\tilde{b}_{2}}\bra{\tilde{b}_{2}}$. Therefore,  we can create an entire disordered lattice with $n$ disordered bonds in a hierarchical fashion with 
\begin{equation}
\textsf{L}^{[n]} = \textsf{L}^{[n-1]} + \ket{\tilde{b}_{n}}\bra{\tilde{b}_{n}}.
\end{equation}
It is clear that each level of this recursion of adding disordered bonds is amenable to an exact inversion via the Sherman Morrison formula in Eq.~\eqref{eq_sherman_morrison}, as enabled by the disordered bonds entering as dyadic additions to the perfect lattice Laplacian. 
We therefore obtain the following recursive relation between the Green's functions $\textsf{G}^{[n]}$ and $\textsf{G}^{[n-1]}$ for a system with $n$ and $n-1$ bond impurities respectively
\begin{equation}
\textsf{G}^{[n]}=\textsf{G}^{[n-1]}+\frac{\textsf{G}^{[n-1]}\ket{\tilde{b}_{n}}\bra{\tilde{b}_{n}}\textsf{G}^{[n-1]}}{1-\mel{\tilde{b}_{n}}{\textsf{G}^{[n-1]}}{\tilde{b}_{n}}}.
    \label{recursion_equation}
\end{equation}
Eq.~\eqref{recursion_equation} represents a central result of our study, with the rest of this section devoted to methods that can yield exact results for these recursion relations. For a small number of disordered bonds these may be computed directly. However, we demonstrate a generalized formalism that enables us to calculate the disordered Green's functions for an arbitrary number of bonds with disorder in an analytically tractable manner.

First, to calculate the Green's function with a single disordered bond, we simply apply the inversion formula in Eq.~\eqref{eq_sherman_morrison} for the Laplacian in Eq.~\eqref{one_bond_laplacian}, which is the $n = 0$ case of the recursion in Eq.~\eqref{recursion_equation}. We thus obtain the  Green's function for a single bond with disorder
\begin{equation}
\label{onebond}
    \textsf{G}^{[1]}=\textsf{G}^{[0]}+\left(\frac{1}{\tilde{g}_1}\right)\textsf{G}^{[0]}\ket{\tilde{b}_1}\bra{\tilde{b}_1}\textsf{G}^{[0]},
\end{equation}
where we have defined 
\begin{subequations}
\label{dressed_green'sfunction_selements}
\begin{align}
\mathcal{\tilde{G}}^0_{\alpha\beta}&\coloneqq \mel{\tilde{b}_{\alpha}}{\textsf{G}^{[0]}}{\tilde{b}_{\beta}}\;\;~~~~ (1\leq \alpha,\beta\leq n)\\
\tilde{g}_\alpha &\coloneqq 1-\mathcal{\tilde{G}}^0_{\alpha\alpha}, \;\; ~~~~~~~~~~~~(1\leq \alpha\leq n) \end{align}
\end{subequations}
We term $\mathcal{\tilde{G}}^0_{\alpha\beta}$ as the disordered {\it dressed} Green's function (as a generalization of the definition in Section \ref{sec:ensembleaverage}). As a limiting case, one can consider the bond percolation limit, where we set the disorder variable $\zeta_1\rightarrow1$ in Eq.~\eqref{onebond} and recover the \textit{single broken bond} Green's function expression derived in \cite{cserti_perturbation, owaidat_infinitesqlattice}, as expected.

Having obtained the Green's function for the system with a single disordered bond, one may obtain the Green's function for two disordered bonds, by using Eq.~\eqref{onebond} in the recursion (Eq.~\eqref{recursion_equation}). On doing so, we obtain the following formula

\begin{align}
 \label{twobond}
&\textsf{G}^{[2]}= \textsf{G}^{[0]}+\left(\frac{\tilde{g}_2}{\tilde{g}_1\tilde{g}_2-(\tilde{\mathcal{G}}_{12}^0)^2}\right)\textsf{G}^{[0]}\ket{\tilde{b}_1}\bra{\tilde{b}_1}\textsf{G}^{[0]}+\\
 \nonumber
&\left(\frac{\tilde{g}_1}{\tilde{g}_1\tilde{g}_2-(\tilde{\mathcal{G}}_{12}^0)^2}\right)\textsf{G}^{[0]}\ket{\tilde{b}_2}\bra{\tilde{b}_2}\textsf{G}^{[0]}+\\
\nonumber
 &\left(\frac{\tilde{\mathcal{G}}_{12}^0}{\tilde{g}_1\tilde{g}_2-(\tilde{\mathcal{G}}_{12}^0)^2}\right)\left[\textsf{G}^{[0]}\ket{\tilde{b}_1}\bra{\tilde{b}_2}\textsf{G}^{[0]}+\textsf{G}^{[0]}\ket{\tilde{b}_2}\bra{\tilde{b}_1}\textsf{G}^{[0]}\right].
\end{align}
Clearly, a pattern is evident for these formulae. The difference between the disordered and perfect lattice Green's function is simply the sum of bilinears of the kind $\textsf{G}^{[0]}\ket{\tilde{b}_\alpha}\bra{\tilde{b}_\beta}\textsf{G}^{[0]}$ for $1\leq \alpha,\beta\leq n$ with disordered coefficients that are functions of the disordered dressed Green's function $\tilde{\mathcal{G}}^{0}_{\alpha\beta}$. Computing these formulae iteratively using the recursion, however, is hard and below we propose an efficient manner in determining the Green's functions. 

Thus, we now account for an arbitrary number of bonds with disorder. We posit the following sum of bilinears as the disordered lattice Green's function with $n$ disordered bonds 
\begin{equation}\label{disordered_green's function_exact}
    \textsf{G}^{[n]}\equiv \textsf{G}^{[0]}+\sum_{\alpha,\beta}^nc_{\alpha\beta}^{[n]}\textsf{G}^{[0]}\ket{\tilde{b}_{\alpha}}\bra{\tilde{b}_{\beta}}\textsf{G}^{[0]}.
\end{equation}
It is straightforward to show that this solves the recursion relation in Eq.~\eqref{recursion_equation}, through mathematical induction. Remarkably, the problem of evaluating the disordered Green's function has now reduced to computing the coefficients $c_{\alpha\beta}^{[n]}$. By symmetry, $c_{\alpha\beta}^{[n]}=c_{\beta\alpha}^{[n]}$ and thus, we need to only compute $n+ n(n+1)/2$ coefficients for $n$ disordered bonds. Multiplying by $(\textsf{G}^{[n]})^{-1}=\left((\textsf{G}^{[0]})^{-1}-\sum_{\alpha}^n\ket{\tilde{b}_\alpha}\bra{\tilde{b}_\alpha}\right)$ on both sides, and using the linear independence of the $\ket{\tilde{b}_\alpha}\bra{\tilde{b}_\beta}$ dyads, we obtain the following relation
\begin{align}\label{generator equation}
    \tilde{g}_\alpha c_{\alpha\beta}^{[n]}-\sum_{\gamma\neq \alpha}\mathcal{\tilde{G}}^0_{\alpha\gamma}c_{\gamma\beta}^{[n]}=\delta_{\alpha\beta} \;\;\; (1\leq \alpha,\beta,\gamma\leq n),
\end{align}
where $\tilde{g}_\alpha$ and $\tilde{\mathcal{G}}^0_{\alpha\gamma}$ are defined as per Eq.~\eqref{dressed_green'sfunction_selements}. The above Eq.~\eqref{generator equation} gives us a linear algebraic relation from which we can determine the coefficients $c^{[n]}_{\alpha\beta}$. The equivalent matrix equation can be formulated in terms of a Kronecker product $\otimes$ as follows
\begin{equation}
    (\tilde{\mathcal{G}}^{[n]}\otimes \mathbb{1}_n)\ket{C^{[n]}}=\ket{U^{[n]}},
    \label{eq_block_structure}
\end{equation}
where  $\ket{C^{[n]}}=(c_{11}^{[n]}  c_{12}^{[n]} \hdots c_{nn}^{[n]})^T$ is the $n^2\times 1$ coefficient vector and $\ket{U^{[n]}}$ is the constant vector with $n$ unit vector blocks, that is $\ket{U^{[n]}}\equiv \ket{e_1^{[n]}|e_2^{[n]}|\hdots |e_n^{[n]}}$ (where  $\ket{e_\alpha^{[n]}}$ denotes the $n$ dimensional $\alpha^{\textrm{th}}$ unit vector). 
The dressed Green's function matrix $\mathcal{\tilde{G}}^{[n]}$ is given explicitly by
\begin{equation}\label{dressed Green's function matrix}
    \mathcal{\tilde{G}}^{[n]}\equiv \begin{pmatrix}
\tilde{g}_1  & -\mathcal{\tilde{G}}^0_{12} & -\mathcal{\tilde{G}}^0_{13} & \cdots & -\mathcal{\tilde{G}}^0_{1n}\\
-\mathcal{\tilde{G}}^0_{21} & \tilde{g}_2  & -\mathcal{\tilde{G}}^0_{23} & \cdots & -\mathcal{\tilde{G}}^0_{2n} \\
\vdots &\vdots &\vdots & \ddots & \vdots \\-\mathcal{\tilde{G}}^0_{n1} &-\mathcal{\tilde{G}}^0_{n2} & -\mathcal{\tilde{G}}^0_{n3} & \cdots & \tilde{g}_n\\
\end{pmatrix},
\end{equation}
The block structure of Eq.~\eqref{eq_block_structure} allows us to solve the matrix equation in a reduced fashion from an $n^2\times n^2$ to an $n\times n$ matrix, which gives us a significant computational advantage. Observe that it suffices to solve $n$ matrix equations of the following kind 
\begin{equation}\label{red_matrix}
    \mathcal{\tilde{G}}^{[n]}\ket{C^{[n]}_\alpha}=\ket{e^{[n]}_\alpha},\\
\end{equation}
where  $\ket{C^{[n]}_\alpha}=(c_{1j}^{[n]}  c_{2\alpha}^{[n]} \hdots c_{n\alpha}^{[n]})^T$ is the $n\times 1$ reduced coefficient vector. Since the right hand side of the matrix equation is simply the unit vector, it is convenient to solve the above systems of equations using Cramer's rule, which gives us the following neat result
\begin{equation}\label{Cramer's rule}
     c_{\alpha\beta}^{[n]}=\frac{\textrm{det}(\mathcal{\tilde{G}}^{[n]}(\alpha,\beta))}{\textrm{det}(\mathcal{\tilde{G}}^{[n]})},
\end{equation}
where, $\mathcal{\tilde{G}}^{[n]}(\alpha,\beta)$ is the coefficient matrix with the $\alpha^{\textrm{th}}$ column replaced with $\ket{e_\beta^{[n]}}$. This simplifies to a particularly provocative form for the diagonal terms as below
\begin{equation}\label{disorered coefficient}
    c_{\alpha\alpha}^{[n]}=\frac{\textrm{det}(\mathcal{G}^{[n-1]})}{\textrm{det}(\mathcal{G}^{[n]})}.
\end{equation}
Here, $\mathcal{G}^{[n]}$ is the coefficient matrix in Eq. \eqref{red_matrix}. Thus, the diagonal terms of the coefficients are simply the ratios of the determinants of the dressed Green's function matrix for $n$ and $n-1$ disordered bonds. This is a remarkable result and suggests that the disordered Green's function are intrinsically determined by the properties of the determinant of a quantity encoding the lattice structure by means of the dressed Green's function matrix defined in Eq.~\eqref{dressed Green's function matrix}. Properties of the coefficients of this matrix should help decipher the response of the system as the disorder strength is increased.

We now enlist the coefficients for the disordered Green's functions for bond impurities of one, two and three bonds respectively. We denote the numerator and denominator of these coefficients by $\mathcal{N}[c_{\alpha\beta}^{[\gamma]}]$ and $\mathcal{D}[c^{[\gamma]}]$ respectively, with
\begin{equation}
c_{\alpha\beta}^{[\gamma]} = \frac{\mathcal{N}[c_{\alpha\beta}^{[\gamma]}]}{\mathcal{D}[c^{[\gamma]}]}.
\end{equation}
Note that due to Eq.~\eqref{Cramer's rule}, there is only one (superscript) index in the coefficient for the denominator since it is equal for all coefficients for a given $n$. Then, for one disordered bond, we have 
\begin{align}\label{onebondcoeff}
    \mathcal{N}[c_{11}^{[1]}]=1; \;\;\; \mathcal{D}[c^{[1]}]=\tilde{g}_1.
\end{align}
for two bond impurities, we have
\begin{subequations}\label{twobondcoeff}
  \begin{align}
      &\mathcal{N}[c_{11}^{[2]}]=\tilde{g}_2;\;\;\mathcal{N}[c_{22}^{[2]}]=\tilde{g}_1; \;\; \mathcal{N}[c_{12}^{[2]}]=\tilde{\mathcal{G}}_{12}^0,\\
     &\mathcal{D}[c^{[2]}]=\tilde{g}_1\tilde{g}_2-(\tilde{\mathcal{G}}_{12}^0)^2 ,
\end{align}
\end{subequations}
and for three bond impurities, we have
\begin{subequations}
\begin{align}
      \mathcal{N}[c_{11}^{[3]}]&=\tilde{g}_2\tilde{g}_3-(\mathcal{\tilde{G}}_{23}^0)^2,\\
      \mathcal{N}[ c_{22}^{[3]}]&=\tilde{g}_1\tilde{g}_3-(\mathcal{\tilde{G}}_{13}^0)^2,\\
        \mathcal{N}[c_{33}^{[3]}]&=\tilde{g}_1\tilde{g}_2-(\mathcal{\tilde{G}}_{12}^0)^2,\\
     \mathcal{N}[c_{12}^{[3]}]&=\tilde{g}_3\tilde{\mathcal{G}}_{12}^0+ \tilde{\mathcal{G}}^0_{13}\tilde{\mathcal{G}}^0_{23},\\
      \mathcal{N}[c_{13}^{[3]}]&=\tilde{g}_2\tilde{\mathcal{G}}_{13}^0+\tilde{\mathcal{G}}^0_{12}\tilde{\mathcal{G}}^0_{23},\\
\mathcal{N}[c_{23}^{[3]}]&=\tilde{g}_1\tilde{\mathcal{G}}_{23}^0+\tilde{\mathcal{G}}^0_{12}\tilde{\mathcal{G}}^0_{31},\\
\nonumber \mathcal{D}[c^{[3]}]&=\tilde{g}_1\tilde{g}_2\tilde{g}_3-\tilde{g}_1(\mathcal{\tilde{G}}_{23}^0)^2-\tilde{g}_2(\mathcal{\tilde{G}}_{31}^0)^2-\tilde{g}_3(\mathcal{\tilde{G}}_{12}^0)^2-\\
    &2\mathcal{\tilde{G}}_{12}^0\mathcal{\tilde{G}}_{23}^0\mathcal{\tilde{G}}_{31}^0.
\end{align}
\end{subequations}
This procedure may be extended to arbitrarily many number of bonds and the coefficients can be obtained as before. Observe that the coefficients in Eqs.~\eqref{onebondcoeff} and \eqref{twobondcoeff} are the same as those computed in  Eqs.~\eqref{onebond} and \eqref{twobond} by directly solving the recursion. Coefficients for four disordered bonds are enlisted in Appendix~\ref{4bond_disorder}.  

\subsection{Connection with perturbation theory}\label{connection with perturbation theory}

The Green's functions computed in this section are exact for arbitrary disorder. To stay consistent, they must match with the predictions from the perturbation theory in the weak disorder regime. This is indeed the case, and we demonstrate here how one can reproduce the perturbation expansion results by simply computing the series expansions of the Green's functions in the disorder variable $\zeta$ in the small disorder strength limit. As an illustration, we consider the Green's function for one disordered bond in Eq.~\eqref{onebond} and computing a series expansion (in $\zeta_1$) of the denominator (under the assumption that $\zeta_1\ll (\mathcal{G}_{11}^{[0]})^{-1}$), we obtain the following expansion,
\begin{align}\label{onebondperturbationexp}
  \nonumber \textsf{G}^{[1]}&=\textsf{G}^{[0]}+\zeta_1   \textsf{G}^{[0]}\ket{b_1}\bra{b_1}\textsf{G}^{[0]}+\zeta_1^2 \mathcal{G}_{11}^{0}\textsf{G}^{[0]}\ket{b_1}\bra{b_1}\textsf{G}^{[0]}+\\
  &\zeta_1^3 (\mathcal{G}_{11}^{0})^2\textsf{G}^{[0]}\ket{b_1}\bra{b_1}\textsf{G}^{[0]}+\hdots.
\end{align}
Now, recall that the perturbation theory predicts the Dyson series (Eq.~\eqref{dyson_series}) for the Green's function. Considering that for one bond disorder, the correction to the Laplacian is simply $\textsf{L}^{(1)}=\zeta_1\ket{b_1}\bra{b_1}$ (as seen from Eq.~\eqref{laplacian_correction dyadic}), on using this in the Dyson series we see that we exactly reproduce Eq.~\eqref{onebondperturbationexp}. This clearly shows how our dyadic bond formulation is consistent with the perturbation theory developed before. This exercise can also be performed for higher number of disordered bonds with the same result.

An alternate benefit of this exercise is that it provides an estimate for the disorder strength upto which the perturbation theory converges. Indeed, we find that as long as $\zeta\ll (\mathcal{G}_{ij}^{[0]})^{-1}$, or equivalently in terms of the disorder strength, $a x_{ij}\ll\ln(1/(1-(\mathcal{G}_{ij}^{[0]})^{-1}))$ where $0\leq x_{ij}\leq 1$ (here, $i$ and $j$ can be identical or nearest-neighbour sites on the lattice), the perturbation theory is expected to provide reasonable results to match with experiments.

Calculating the system variables from these exact Green's functions is straightforward---$\ket{V}=\textsf{G}^{[n]}\ket{I}$, and $\ket{J_{\bhat{e}}}=-\textsf{D}^{(0)}_{\bhat{e}}\ket{V}$. Further, just as we did in the perturbation theory, these disordered Green's functions can be averaged analytically to obtain the ensemble-averaged nodal voltages and bond currents. We expand on this in greater detail in subsection \ref{nodal voltage fluctuations}. 

\begin{figure*}[t!]
    \includegraphics[width=\textwidth]{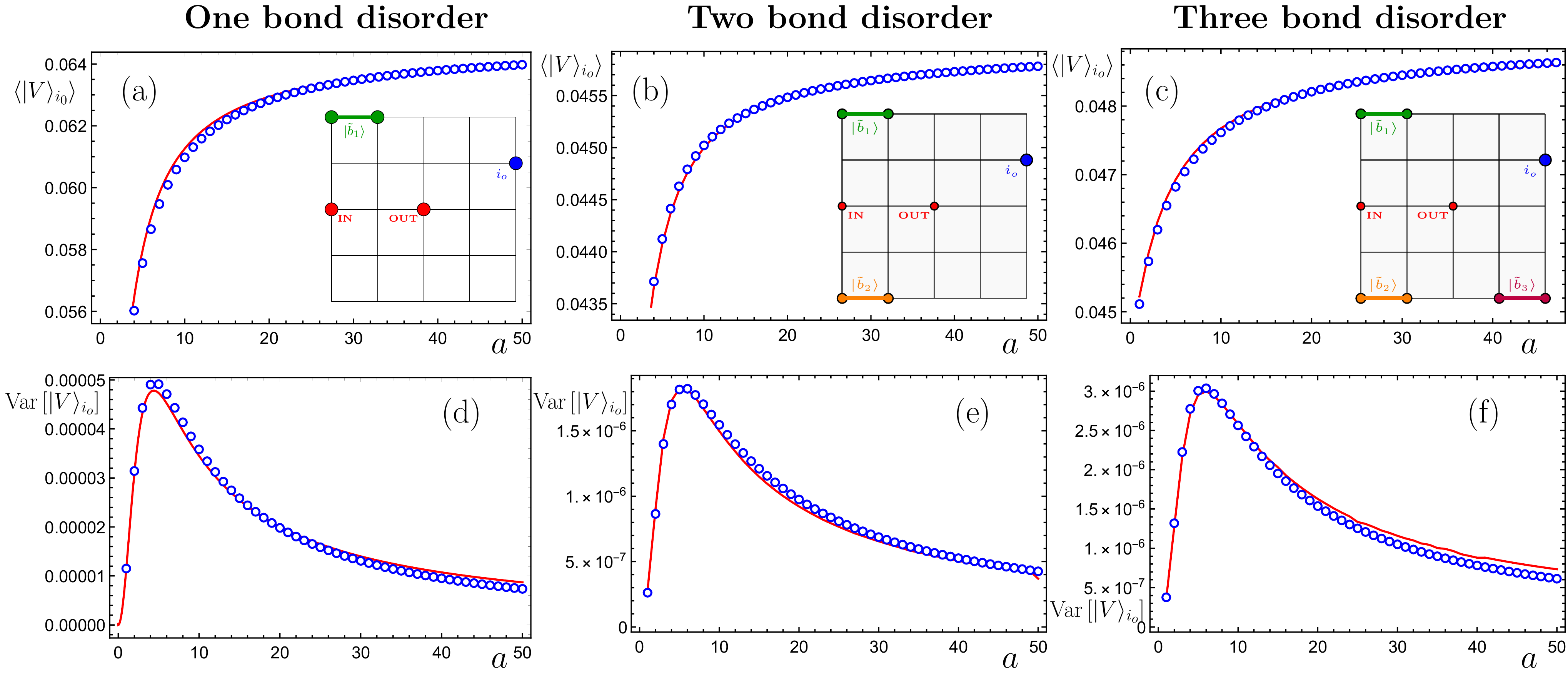}
    \caption{Plot of the disorder-averaged nodal voltages $\langle \ket{V}_{i_o}\rangle $ in subfigures $\{(a), (b), (c)\}$ and their fluctuations $\textrm{Var}\left[\ket{V}_{i_o}\right]$ in subfigures $\{(d), (e), (f)\}$ for one, two and three disordered bonds in the lattice respectively. The plots display an exact match between the theory (red line) and numerical simulations (blue circles). The insets display a schematic of the $5\times 5$ RRN used in the computation, with the red circles denoting the locations of the source and sink, whereas the green, orange and purple bonds depicting the disordered bonds $\ket
    {\tilde{b}_1}, \ket{\tilde{b}_2}$ and $\ket{\tilde{b}_3}$ respectively, and the blue circle denotes the observation point $i_o$. The averaging is performed over $1000$ realizations of the bond disorder. The analytic predictions were computed using Eqs.~\eqref{nodal_voltage_average} and~\eqref{second_moment_nodal_voltage}, with the exact formulae for the $n$-dimensional integrals therein given in Appendix~\ref{dis_coefficient_formulae}.} 
   \label{fig:voltage fluct}
\end{figure*}

\subsection{Numerical Implementation}\label{Numerical Implementation}
At this point, it is useful to discuss the applicability of the recursive method from a numerical standpoint. To compute a disordered Green's function $\textsf{G}$, assuming that the perfect lattice Green's function $\textsf{G}^{[0]}$ is known, one typically inverts the disordered Laplacian $\textsf{L}$. Since the Laplacian is a matrix of size $N_s\times N_s$, this operation costs $O((N_s)^3)$ time. We argue that using the recursive formula in Eq.~\eqref{recursion_equation} can provide a computational advantage against direct inversion. In the recursion, one needs to compute outer products of the kind $\textsf{G}^{[n-1]}\ket{b_\alpha}\bra{b_\beta}\textsf{G}^{[n-1]}$. Naively, computing $\textsf{G}^{[n-1]}\ket{b_\alpha}$ costs $O((N_s)^2)$ time, however, we note that $\ket{b_\alpha}=\ket{s_i}-\ket{s_j}$ (where $\alpha$ is the index for the bond between $i$ and $j$), and since the site vectors are unit vectors, such a computation is equivalent to accessing a column from the Green's function matrix, which can be performed in $O(N_s)$ time. Thus, computing the outer product requires $O((N_s)^2)$ time, and multiplying it with a disorder variable $\zeta_\alpha$ and coefficient $(1/(1-\mel{b_\alpha}{G^{[n-1]}}{b_\beta}))$ can both be performed in $O((N_s)^2)$ time (as well as adding the result to $\textsf{G}^{[n-1]}$). Now, we note that for $n$ disordered bonds, the recursion runs $n$ times and the total time complexity becomes $O(n(N_s)^2)$ time. This becomes equivalent to the complexity for linear inversion for disorder in all bonds $(n=N_b=O(N_s))$, but is \textit{faster} whenever $n$ is sublinear in $N_b$---an advantage which is not accessible in the linear inversion technique. Thus, our recursive method also provides a numerically viable approach to compute the lattice Green's function for a disordered network. It would be interesting to devise algorithms to test such an implementation and compare efficiency with other recursive techniques such as the transfer matrix approach \cite{derrida1982transfer}, which we have not attempted here.

In addition to the recursive algorithm, we also provide the exact solution to the recursion in Eqs.~\eqref{disordered_green's function_exact} and ~\eqref{Cramer's rule}. While they provide excellent analytical control, from a numerical standpoint, computation of the coefficients may appear to be expensive due to the evaluation of the determinants in Eq.~\eqref{Cramer's rule}. Here, we point out instead of computing $O(n^2)$ many determinants, one only needs to compute three---the denominator $\mathcal{D}[c^{[n]}]$, and two numerators $\mathcal{N}[c^{[n]}_{\alpha\alpha}]$ and $\mathcal{N}[c^{[n]}_{\alpha\beta}]$ where $\alpha,\beta$ can be any index from $1$ to $n$. The rest of the determinants can be constructed by inspection (since they have the same polynomial form, for instance, see Appendix~\ref{4bond_disorder}). 
It is, therefore, reasonable to expect that a symbolic computation of this expression in the disorder variables $\{\zeta_{ij}\}$ can enable a disorder average for the Green's function in a much more efficient manner than inverting a Laplacian each time. Thus, there does appear to be a significant advantage in devising algorithms using the recursive method.

\section{Numerical results}\label{sec:numerics}

Finally, we test the validity of our theoretical results through numerical simulation of an $L \times L$ square lattice in two dimensions with exponential disorder. In subsection~\ref{nodal voltage fluctuations}, we first numerically compute the mean nodal voltages and its fluctuations at a generic lattice site for one, two, and three disordered bonds in the lattice. These results are then matched with theoretical predictions from the hierarchical dyadic bond disorder formalism, which show an exact match for the entire disorder range for the weak as well as strong regimes. In subsection~\ref{bond current fidelity}, we propose a novel order parameter, termed \textit{bond current fidelity}, which measures the overlap between current distributions at arbitrary and infinite disorder. We provide finite-size scaling estimates of this order parameter in the weak and strong disorder regime, albeit for small lattice sizes.

\subsection{Nodal voltage fluctuations}\label{nodal voltage fluctuations}
We first numerically compute the nodal voltages and fluctuations at an observation site $i_o$ in the lattice. The nodal voltages $\{\ket{V}_{i_o}\}$ are calculated using Kirchhoff's law (Eq.~\eqref{lvi}) while the fluctuations are given simply as an average over disorder realizations $\textrm{Var}[\ket{V}_{i_o}]\equiv \langle \ket{V}_{i_o}^2  \rangle-\langle \ket{V}_{i_o}  \rangle^2$. The choice of the observation site $i_o$ is arbitrary, as our formalism predicts the values at all lattice sites. We notice and expect a similar behaviour at any such lattice site in the bulk and thus it suffices to probe the behaviour at a single arbitrary lattice site. For illustrative purposes, our simulations are performed on a $5 \times 5$ square lattice with one, two and three disordered bonds. The locations of the disordered bonds and the observation site are depicted in the schematic in the insets of Fig.~\ref{fig:voltage fluct}(a), (c) and (e).

For the cases with one, two and three disordered bonds, one can also compute analytically the quantities of interest using the recursive dyadic bond disorder formulation. It is easy to compute the moments of the nodal voltages using Eq.~\eqref{disordered_green's function_exact}. The mean nodal voltage at site $i_o$ is given by
\begin{equation}\label{nodal_voltage_average}
\left\langle\ket{V}_{i_o}\right\rangle=\ket{V}^{(0)}_{i_o}+\sum_{\alpha,\beta}^n\mathcal{I}_{\alpha\beta}^{[n]}(a)\:\mel{s_{i_o}}{\textsf{G}^{[0]}}{b_{\alpha}}\mel{b_{\beta}}{\textsf{G}^{[0]}}{I},
\end{equation}
where $\mathcal{I}_{\alpha\beta}^{[n]}(a)$ is a disorder averaged coefficient obtained by computing an $n$-dimensional integral over the disorder distributions (see Appendix \ref{dis_coefficient_formulae} for explicit formulae). These integrals may be computed analytically for simple cases (and numerically otherwise) and the computed nodal voltages are matched with the numerics (cf. Fig.~\ref{fig:voltage fluct}(a), (c) and (e)). We can also compute the fluctuations by a somewhat more tedious calculation. For this, we must compute the second moment of the nodal voltage at site $i_{o}$, which may be written as 
\begin{align}\label{second_moment_nodal_voltage}
\nonumber&\left\langle\left(\ket{V}_{i_o}\right)^2\right\rangle=\left(\ket{V}_{i_o}^{(0)}\right)^2+\\
\nonumber &2\ket{V}_{i_o}^{(0)} \sum_{\alpha,\beta}^n\mathcal{I}_{\alpha\beta}^{[n]}\:\mel{s_{i_o}}{\textsf{G}^{[0]}}{b_{\alpha}}\mel{b_{\beta}}{\textsf{G}^{[0]}}{I}+\\
&\sum_{\alpha,\beta,\gamma,\delta}^n\!\!\!\mathcal{J}_{\alpha\beta\gamma\delta}^{[n]}\mel{s_{i_o}}{\textsf{G}^{[0]}}{b_{\alpha}}\!\mel{s_{i_o}}{\textsf{G}^{[0]}}{b_{\beta}}\!\mel{b_{\gamma}}{\textsf{G}^{[0]}}{I}\!\mel{b_{\delta}}{\textsf{G}^{[0]}}{I},
\end{align}
where $\mathcal{I}_{\alpha\beta}^{[n]
}(a)$ and $\mathcal{J}_{\alpha\beta\gamma\delta}^{[n]}(a)$ are disorder averaged coefficients. $\mathcal{I}_{\alpha\beta}^{[n]}(a)$ is the $n$-dimensional integral described above, while $\mathcal{J}_{\alpha\beta\gamma\delta}^{[n]}(a)$ is an $n$-dimensional integral connecting four bonds (again, exact formulae are provided in Appendix~\ref{dis_coefficient_formulae}). Clearly, the fluctuations are given simply by the four-point term (the third term in Eq.~\eqref{second_moment_nodal_voltage}) with a disorder averaged coefficient given by $\mathcal{J}_{\alpha\beta}^{[n]}(a)-\mathcal{I}_{\alpha\beta}^{[n]}(a)\mathcal{I}_{\gamma\delta}^{[n]}(a)$. We show the match of the voltage fluctuations with the simulations in Fig.~\ref{fig:voltage fluct}(b), (d) and (f). 
From Fig.~\ref{fig:voltage fluct}, we notice that the nodal voltages increase with increasing disorder strength $a$, and then saturate to a constant value. The saturation is due to the fact that in the strong disorder limit, the current distribution collapses to an optimal path, and hence small changes in the disorder strength does not alter the voltage configuration of the network substantively. 
The fluctuations, on the other hand, depict a very interesting behaviour---they peak at a critical disorder value and then decrease as we approach the strong disorder limit. This peak in the fluctuations is representative of the crossover between the weak and strong disorder regimes in the system, for the chosen configuration of small number of impurities. The fluctuations in the nodal voltages thus may be considered as a useful order parameter for the system, at least, for systems with a small number of bonds with disorder. 

Although, we have provided theoretical expressions and numerical results for a small lattice size, the case of an infinite lattice size with one, two and three disordered bonds does not require greater effort, as the only change is the replacement of the Green's function elements for the infinite lattice. Analytic expressions for the same are easy to compute through known recursion relations for the perfect infinite lattice Green's function~\cite{cserti_greensfunction_ajp, katsura_green}.

\subsection{Bond current fidelity}\label{bond current fidelity}

To investigate the behaviour of the system in the weak and strong disorder regimes, we construct a novel order parameter in terms of the bond current observables. Previous studies of the crossover between weak and strong disorder regimes have focused on microscopic observables such as the distribution of tracer path lengths~\cite{stanley_optimalpaths}, which requires collecting large statistics over many samples, or through the resistance measurements before and after introduction of a perturbation at the bond with a maximal current~\cite{percolationtransition_afrydman}. In this study, we propose an alternate order parameter computed using a macroscopic observable---the bond currents, thus making the study accessible to experiments. The advantage of this parameter is that it solely depends on the steady state current distribution of a single resistance configuration, and thus does not require, in principle, large averaging, or perturbations to the circuit. Knowledge of the resistance configuration, from an experimental point-of-view requires the simultaneous measurement of the nodal voltages and bond currents, which appears to be accessible within reasonable arrangements. 

In order to quantify whether a given system is in a weak or strong disordered phase, it is useful to look at how \textit{channelized} the flow of the current is. Each resistance configuration has a corresponding optimal path, that is, the path of least resistance between the source and the sink in the circuit. For each disorder strength, the optimal path can be computed by simply computing the path of least total weight in the graph between the source and sink, which is implemented efficiently by Djikstra's algorithm~\cite{djikstra}. In the perfect lattice, the optimal path is the straight line connecting the source and the sink. As the disorder is increased, in the weak regime, the optimal path fluctuates about this straight line, with its statistics depicting a self-affine behaviour~\cite{banavar_optimalpaths}. As the system enters the strong disorder regime, the optimal path changes behaviour--- displaying self-similarity. We recover these signatures from the bond current order parameter described in detail below.

In the strong disorder limit, the current distribution collapses completely to the optimal path. Therefore, a relevant order parameter is the {\it deviation} of the current distribution in the system from the optimal path. To define the order parameter, it is useful to define the optimal path in terms of bond vectors. Similar to the definition of the bond current vector in Sec.~\ref{sec:rrn}, we construct $d$ $N_s$-dimensional optimal current vectors $\{\ket{J^{\textrm{opt}}_{\bhat{e}}(a)}\}$ along $d$ directions and an $N_b$-dimensional complete optimal current vector $\ket{J^{\textrm{opt}}(a)}\equiv\ket{J^{\textrm{opt}}_{\bhat{e}_1}|J^{\textrm{opt}}_{\bhat{e}_2}|\hdots |J^{\textrm{opt}}_{\bhat{e}_d}}$. We define $\ket{J^{\textrm{opt}}_{\bhat{e}}(a)}$ as follows
\begin{equation}
     \ket{J^{\textrm{opt}}_{\bhat{e}}(a)}_{i}\coloneqq \begin{cases}1 & \textrm{if} \;\; \ket{b_\alpha} \in \textrm{optimal path} \\
     &\textrm{and}\:\:\: \alpha\equiv {(i,\bhat{e})} \\
    -1 &\textrm{if} \;\; \ket{b_\alpha} \in \textrm{optimal path} \\ &\textrm{and} \:\: \alpha\equiv (j,-\bhat{e})\:\: \textrm{with}\:\: \ijplus \\
    0&  \textrm{otherwise}\end{cases}.
\end{equation}
This construction is consistent with the current definitions and accounts for possible overhangs in the optimal path. Note that the length of the optimal path is given by the 1-norm of the optimal current vector. We then define the bond current fidelity $\mathcal{F}$ to be given by
\begin{equation}
    \mathcal{F}_{\{\zeta\}}(a)\equiv \bra{J^{\textrm{opt}}(a)}\ket{J(a)},
\end{equation}
where the subscript $\{\zeta\}$ indicates that the fidelity is calculated for a fixed instance of the resistance disorder configuration, with the only variable being the disorder strength that is varied. We conjecture that this order parameter $ \mathcal{F}$ should measure the behaviour of the system succinctly with respect to the behaviour of the optimal path, and one should notice a significant shift in the profile from the self-affine to self-similar profiles of the optimal path in the weak and strong disorder regimes respectively. 
Since this parameter is an \textit{overlap} between the actual current distribution and the optimal path at each disorder strength, we term it the \textit{bond current fidelity} of the system. Similar fidelity parameters have been found useful in a variety of contexts, including in information geometric contexts for probing quantum phase transitions~\cite{fidelity_gu}. 

\begin{figure}[t!]
    \includegraphics[width=\columnwidth]{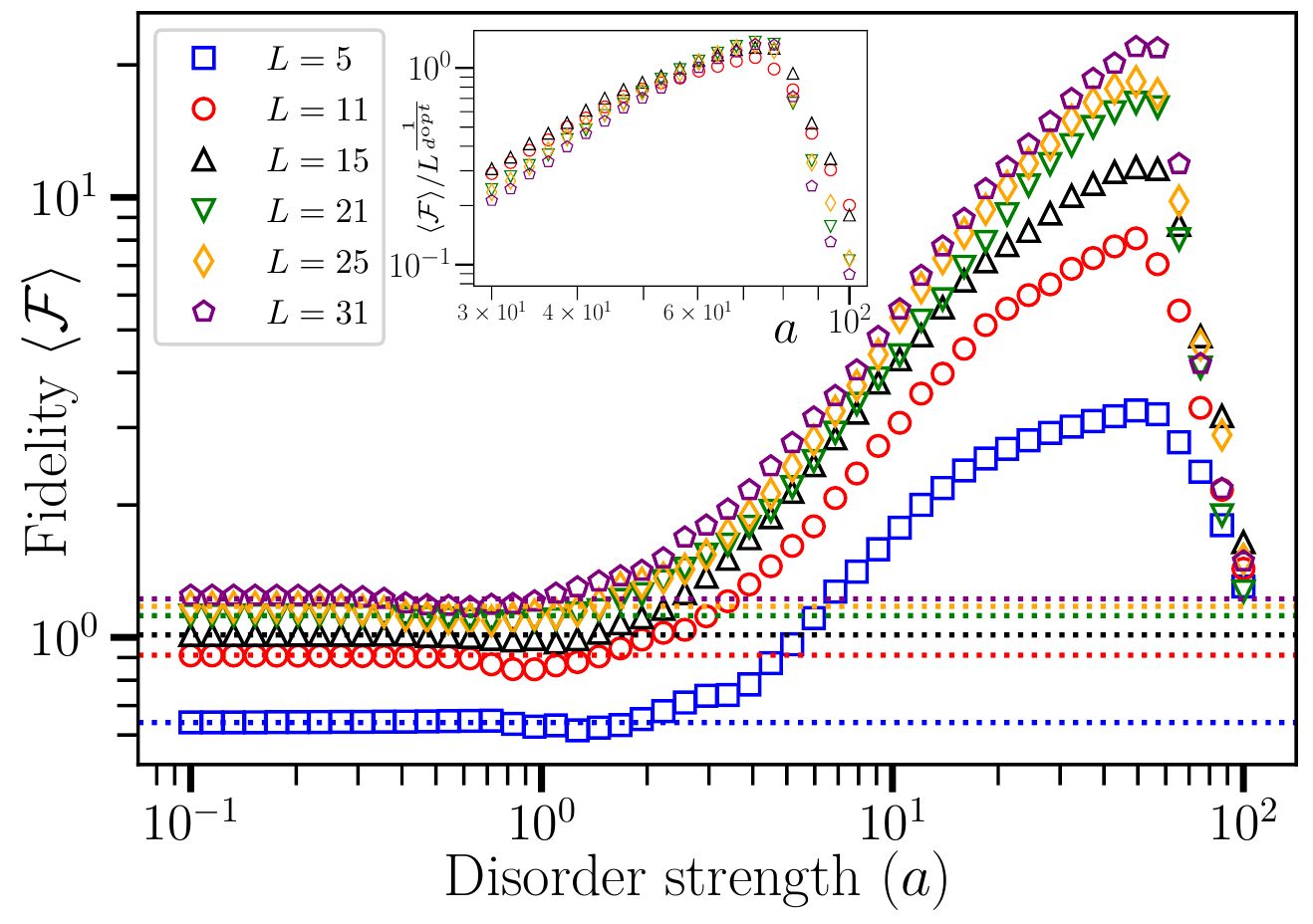}
    \caption{Double logarithm plot of the averaged bond current fidelity $\langle \mathcal{F}\rangle$ calculated for six system sizes $L=5, 11, 15, 21, 25$ and $31$. The strength of disorder is $0.1\leq a\leq 100$. The averaging is performed over 50 realizations of the resistance configuration. The horizontal dotted lines show the saturation values obtained analytically using Eq.~\eqref{fweak}, that is, $\mathcal{F}(a=0)=0.64, 0.91, 1.01, 1.12, 1.17, 1.22$ for the six system sizes respectively. Inset: Scaling collapse for the averaged bond current fidelity in a double logarithm plot, with the symbols denoting five system sizes as per the legend in the main plot, from $L=11$ to $L=31.$ The averaged bond fidelity in the strong disorder regime is scaled by $L^{1/d^{\textrm{opt}}}$ where $d^{\textrm{opt}}=1.22$ is the optimal path exponent in two dimensions.}
   \label{fig:fidelity}
\end{figure}

In Fig.~\ref{fig:fidelity}, we show numerical simulations of the averaged  bond current fidelity for six different system sizes $L = 5,11,15,21,25$ and $31$ given a fixed disorder configuration. The disorder strength is varied from the weak to the strong disorder regimes, that is, $0.1\leq a\leq 100$. We observe interesting signatures of the weak and strong disorder regimes in this order parameter. In the weak disorder regime ($L\gg a^{4/3}$) we observe a saturated behaviour of the fidelity, which is very close to the value of the fidelity in the perfect network, with negligible fluctuations. For our largest system size ($L=31$), we find that weak disorder is given at $a\ll 13$, while strong disorder is given at $a\gg 13$. This is consistent with the profiles observed in Fig.~\ref{fig:fidelity}. We can also compute the saturation value of the weak disorder limit explicitly using the following formula
\begin{equation}\label{fweak}
\mathcal{F}(a=0)=  \sum_{i_{\textrm{in}}}^{i_{\textrm{out}}}\ket{J_{\bhat{e}_x}^{(0)}}_{i}=
\sum_{i_{\textrm{in}}}^{i_\textrm{out}}\mel{s_{i}}{\textsf{D}^{(0)}_{\bhat{e}_x}\textsf{G}^{(0)}}{I},
\end{equation}
where the bond currents have been computed using the lattice Green's function as per Eq.~\eqref{dvj}. These values are depicted in Fig.~\ref{fig:fidelity} by the horizontal dotted values, explicitly given by $\mathcal{F}(a=0)=0.64, 0.91, 1.01, 1.12, 1.17, 1.22$ for system sizes $L=5,11,15,21,25$ and $31$ respectively. We further note that as described in Eq.~\eqref{dvj}, the currents are computed using the Green's functions, which scale logarithmically with system size in two dimensions. This is consistent with the aforementioned saturation values obtained from explicit computation which scale as $\log(L)$ with increasing system size $L$.   

In the strong disorder regime ($L\ll a^{4/3}$), we notice a steady rise in the fidelity upto a critical disorder strength, after which there is a rapid fall in the fidelity to a value that is simply given by the length of the optimal path at infinite disorder. This critical disorder appears to be the state of the system with maximum fluctuations from optimal behaviour. As previously understood from literature, the scaling of the system in the strong disorder limit are given by the optimal path exponent $d^{\textrm{opt}}=1.22$. Likewise, we find a scaling collapse of the system in the strong disorder regime using this exponent, in particular, the collapse is obtained by scaling the fidelities by $L^{1/d^{\textrm{opt}}}$.  Finally, an additional curious observation is a systematic decrease in the bond current fidelity in the crossover regime between the weak and strong disorders, which, in fact, is also the global minimum of the fidelity for all system sizes. This appears to be a feature of the crossover regime which demands further investigation.

\section{Discussion and conclusion}\label{sec:discussion}

In this paper, we have demonstrated a framework to determine the disordered Green's functions for RRNs. We investigate two formulations, a perturbative framework that can be used to compute system properties in the weak disordered regime, and a novel exact (dyadic) framework that can be used to compute quantities at even arbitrarily large disorder strengths. We demonstrated the equivalence of the two frameworks in the weak disorder regime. In addition, we obtained explicit Green's functions for arbitrary disorder strength---with analytically tractable expressions---for lattices with small number of disordered bonds. We also demonstrated an exact match between numerical and analytical predictions for nodal voltage fluctuations in such lattices with small numbers of impurities (one, two and three). We have also explicitly demonstrated the generalization of the dyadic framework for the case with a large number of disordered bonds in the system, however, in this case, the explicit computations become more tedious to perform. Finally, we proposed a novel order parameter, named the bond current fidelity, which measures the deviation of the currents from the optimal path given a disorder configuration. We found that this order parameter is able to distinguish between the weak and strong disorder regimes of the system, and finite size scaling estimates are consistent with behaviour obtained in previous studies.

There are several directions for further investigation that our study suggests. In this study, we have focused on fundamental observables relevant to experiments, i.e. nodal voltages and bond currents, and our theoretical predictions for the ensemble averages would be interesting to match with steady-state experimental observations. As RRNs with exponential disorder are paradigmatic systems in condensed matter and statistical physics~\cite{halperin_RRN, doussal, stanley_pre, percolationtransition_afrydman, frydman_percolationmodel}, our results have interesting implications for the characteristics of systems with wide disorder distributions. In particular, it would be very interesting if our dyadic bond formulation can be systematically computed in a manner that sheds more light on the optimal path exponent ($\approx 1.22$ in two dimensions). In this endeavour, it appears that the statistics of the determinant in Eq.~\eqref{disorered coefficient} are of crucial importance, and a systematic study of the growth of these determinants could lead to an understanding of the behaviour of optimal paths in such systems.

We also draw an interesting connection of our work to the paradigmatic model of wave localization in condensed matter systems---Anderson localization. While the current localization we observe is a \textit{classical} and \textit{non-wave} phenomena---the localization only occurs due to the geometric properties of the network, like in a classical percolation model---we note that a \textit{mathematical} correspondence can be drawn between the lattice Green's function and a tight-binding Hamiltonian \cite{murphyprl, imryprx, movassagh2017green}. Thus, we suspect that properties of disordered Green's functions (as analytically computed through our perturbative and exact methods) could explain eigenvector localization in Anderson Hamitlonians with bond-disorder (also known as off-diagonal disorder \cite{ziman1982localization}).

Further, the results presented in this paper are easily generalizable to any underlying disorder, not just exponential, and it would be interesting to study how the nature of the underlying disorder changes the behaviour studied here. In addition, it should be possible to use the scheme of constructing a disordered Green's function as described in Sec.~\ref{sec:dyadicbond}, to construct any arbitrary electrical circuit, such as a bipartite lattice with two types of bonds, a context frequently relevant in condensed matter systems. There are also intriguing connections to be made to the eigenvalues of disordered Laplacians (which could represent Hessian matrices) through the study of their resolvents, which are fundamental quantities of interest in stability studies of random media. 

Finally, in the steady-state regime of the network, we studied the moments of the voltages and bond currents and demonstrated techniques to compute them at arbitrary orders exactly. However, given the addition of another fluctuating degree of freedom (for instance, in studying transient properties of such networks), one could investigate the differences between quenched and annealed disorder by choosing to perform the disorder averaging at the level of the moment-generating function or the cumulant generating function respectively. In particular, quenched disorder may also require the need to use the replica trick \cite{edwards1975theory}. Thus, it would be interesting to check if the methods developed in this paper can controllably demonstrate noteworthy  differences between quenched and annealed disorder in the approach to steady-state of such disordered systems.

\subsection*{Acknowledgments}
We acknowledge useful discussions with Debankur Das, Roshan Maharana, Soham Mukhopadhyay and Shamashis Sengupta. SB also acknowledges useful discussions with Adhip Agarwala and Harish Adsule, and support from the KVPY fellowship. We also thank the anonymous referees for their valuable comments. This project was funded by intramural funds at TIFR Hyderabad from the Department of Atomic Energy (DAE), Government of India.

\appendix

\section{Exact Green's functions for four disordered bonds}\label{4bond_disorder}

Here, we present exact disordered Green's functions for four disordered bonds. For this case we have the following quartic polynomial in $\tilde{g}$ as the denominator of the coefficients:
\begin{align}
    \nonumber &\mathcal{D}[c^{[4]}]=\tilde{g}_1\tilde{g}_2\tilde{g}_3\tilde{g}_4-\\
    \nonumber &\tilde{g}_3\tilde{g}_4(\tilde{\mathcal{G}}_{12}^{0})^2-\tilde{g}_2\tilde{g}_4(\tilde{\mathcal{G}}_{13}^{0})^2-\tilde{g}_2\tilde{g}_3(\tilde{\mathcal{G}}_{14}^{0})^2-\tilde{g}_1\tilde{g}_4(\tilde{\mathcal{G}}_{23}^{0})^2-\\
    \nonumber &\tilde{g}_1\tilde{g}_3(\tilde{\mathcal{G}}_{24}^{0})^2-\tilde{g}_1\tilde{g}_2(\tilde{\mathcal{G}}_{34}^{0})^2-2(\tilde{g}_4\tilde{\mathcal{G}}_{12}^{0}\tilde{\mathcal{G}}_{13}^{0}\tilde{\mathcal{G}}_{23}^{0}+\\
    \nonumber &\tilde{g}_3\tilde{\mathcal{G}}_{12}^{0}\tilde{\mathcal{G}}_{14}^{0}\tilde{\mathcal{G}}_{24}^{0}+\tilde{g}_2\tilde{\mathcal{G}}_{13}^{0}\tilde{\mathcal{G}}_{14}^{0}\tilde{\mathcal{G}}_{34}^{0}+\tilde{g}_1\tilde{\mathcal{G}}_{23}^{0}\tilde{\mathcal{G}}_{24}^{0}\tilde{\mathcal{G}}_{34}^{0})+\\
    \nonumber &(\tilde{\mathcal{G}}_{12}^{0})^2(\tilde{\mathcal{G}}_{34}^{0})^2+(\tilde{\mathcal{G}}_{13}^{0})^2(\tilde{\mathcal{G}}_{24}^{0})^2+(\tilde{\mathcal{G}}_{14}^{0})^2(\tilde{\mathcal{G}}_{23}^{0})^2-\\
    &2(\tilde{\mathcal{G}}_{12}^{0}\tilde{\mathcal{G}}_{13}^{0}\tilde{\mathcal{G}}_{24}^{0}\tilde{\mathcal{G}}_{34}^{0}+\tilde{\mathcal{G}}_{12}^{0}\tilde{\mathcal{G}}_{14}^{0}\tilde{\mathcal{G}}_{23}^{0}\tilde{\mathcal{G}}_{34}^{0}+\tilde{\mathcal{G}}_{13}^{0}\tilde{\mathcal{G}}_{14}^{0}\tilde{\mathcal{G}}_{23}^{0}\tilde{\mathcal{G}}_{24}^{0})
\end{align}
We also compute the numerators of the coefficients, which are cubic polynomials in $\tilde{g}$. Thus, we have
\begin{align}
 \nonumber \mathcal{N}[c_{11}^{[4]}]&=\tilde{g}_2\tilde{g}_3\tilde{g}_4-\tilde{g}_4(\mathcal{\tilde{G}}_{23}^0)^2-\tilde{g}_3(\mathcal{\tilde{G}}_{24}^0)^2-\tilde{g}_2(\mathcal{\tilde{G}}_{34}^0)^2-\\
    &2\mathcal{\tilde{G}}_{23}^0\mathcal{\tilde{G}}_{24}^0\mathcal{\tilde{G}}_{34}^0\\
\nonumber \mathcal{N}[c_{22}^{[4]}]&=\tilde{g}_1\tilde{g}_3\tilde{g}_4-\tilde{g}_4(\mathcal{\tilde{G}}_{13}^0)^2-\tilde{g}_3(\mathcal{\tilde{G}}_{14}^0)^2-\tilde{g}_1(\mathcal{\tilde{G}}_{34}^0)^2-\\
    &2\mathcal{\tilde{G}}_{13}^0\mathcal{\tilde{G}}_{14}^0\mathcal{\tilde{G}}_{34}^0\\
\nonumber \mathcal{N}[c_{33}^{[4]}]&=\tilde{g}_1\tilde{g}_2\tilde{g}_4-\tilde{g}_4(\tilde{\mathcal{G}}_{12}^0)^2-\tilde{g}_2(\mathcal{\tilde{G}}_{14}^0)^2-\tilde{g}_1(\mathcal{\tilde{G}}_{24}^0)^2-\\
    &2\mathcal{\tilde{G}}_{12}^0\mathcal{\tilde{G}}_{14}^0\mathcal{\tilde{G}}_{24}^0\\
\nonumber \mathcal{N}[c_{44}^{[4]}]&=\tilde{g}_1\tilde{g}_2\tilde{g}_3-\tilde{g}_3(\mathcal{\tilde{G}}_{12}^0)^2-\tilde{g}_2(\mathcal{\tilde{G}}_{13}^0)^2-\tilde{g}_1(\mathcal{\tilde{G}}_{23}^0)^2-\\
&2\mathcal{\tilde{G}}_{12}^0\mathcal{\tilde{G}}_{13}^0\mathcal{\tilde{G}}_{23}^0\\
\nonumber\mathcal{N}[c_{12}^{[4]}]&=\tilde{g}_3\tilde{g}_4\mathcal{\tilde{G}}_{12}^0+\tilde{g}_4\tilde{\mathcal{G}}_{13}^0\tilde{\mathcal{G}}_{23}^0+\tilde{g}_3\mathcal{\tilde{G}}_{14}^0\mathcal{\tilde{G}}_{24}^0+\\
     &\tilde{\mathcal{G}}_{13}^0\tilde{\mathcal{G}}_{24}^0\tilde{\mathcal{G}}_{34}^0+\tilde{\mathcal{G}}_{14}^0\tilde{\mathcal{G}}_{23}^0\tilde{\mathcal{G}}_{34}^0-(\tilde{\mathcal{G}}_{12}^0)^2\tilde{\mathcal{G}}_{34}^0\\
\nonumber\mathcal{N}[c_{13}^{[4]}]&=\tilde{g}_2\tilde{g}_4\tilde{\mathcal{G}}_{13}^0+\tilde{g}_4\tilde{\mathcal{G}}_{12}^0\tilde{\mathcal{G}}_{23}^0+\tilde{g}_2\tilde{\mathcal{G}}_{14}^0\tilde{\mathcal{G}}_{34}^0+\\
     &\tilde{\mathcal{G}}_{12}^0\tilde{\mathcal{G}}_{24}^0\tilde{\mathcal{G}}_{34}^0+\tilde{\mathcal{G}}_{14}^0\tilde{\mathcal{G}}_{23}^0\tilde{\mathcal{G}}_{34}^0-(\tilde{\mathcal{G}}_{13}^0)^2\tilde{\mathcal{G}}_{24}^0\\ 
\nonumber\mathcal{N}[c_{14}^{[4]}]&=\tilde{g}_2\tilde{g}_3\tilde{\mathcal{G}}_{14}^0+\tilde{g}_3\tilde{\mathcal{G}}_{12}^0\tilde{\mathcal{G}}_{24}^0+\tilde{g}_2\tilde{\mathcal{G}}_{13}^0\tilde{\mathcal{G}}_{34}^0+\\
     &\tilde{\mathcal{G}}_{12}^0\tilde{\mathcal{G}}_{23}^0\tilde{\mathcal{G}}_{34}^0+\tilde{\mathcal{G}}_{13}^0\tilde{\mathcal{G}}_{23}^0\tilde{\mathcal{G}}_{24}^0-(\tilde{\mathcal{G}}_{14}^0)^2\tilde{\mathcal{G}}_{23}^0\\
\nonumber\mathcal{N}[c_{23}^{[4]}]&=\tilde{g}_1\tilde{g}_4\tilde{\mathcal{G}}_{23}^0+\tilde{g}_4\tilde{\mathcal{G}}_{12}^0\tilde{\mathcal{G}}_{13}^0+\tilde{g}_1\tilde{\mathcal{G}}_{24}^0\tilde{\mathcal{G}}_{34}^0+\\
    &\tilde{\mathcal{G}}_{12}^0\tilde{\mathcal{G}}_{14}^0\tilde{\mathcal{G}}_{34}^0+\tilde{\mathcal{G}}_{13}^0\tilde{\mathcal{G}}_{14}^0\tilde{\mathcal{G}}_{24}^0-(\tilde{\mathcal{G}}_{14}^0)^2\tilde{\mathcal{G}}_{23}^0\\
\nonumber\mathcal{N}[c_{24}^{[4]}]&=\tilde{g}_1\tilde{g}_3\tilde{\mathcal{G}}_{24}^0+\tilde{g}_3\tilde{\mathcal{G}}_{12}^0\tilde{\mathcal{G}}_{14}^0+\tilde{g}_1\tilde{\mathcal{G}}_{23}^0\tilde{\mathcal{G}}_{34}^0+\\
    &\tilde{\mathcal{G}}_{12}^0\tilde{\mathcal{G}}_{13}^0\tilde{\mathcal{G}}_{34}^0+\tilde{\mathcal{G}}_{13}^0\tilde{\mathcal{G}}_{14}^0\tilde{\mathcal{G}}_{23}^0-(\tilde{\mathcal{G}}_{13}^0)^2\tilde{\mathcal{G}}_{24}^0\\ 
\nonumber\mathcal{N}[c_{34}^{[4]}]&=\tilde{g}_1\tilde{g}_2\tilde{\mathcal{G}}_{34}^0+\tilde{g}_2\tilde{\mathcal{G}}_{13}^0\tilde{\mathcal{G}}_{14}^0+\tilde{g}_1\tilde{\mathcal{G}}_{23}^0\tilde{\mathcal{G}}_{24}^0+\\
     &\tilde{\mathcal{G}}_{12}^0\tilde{\mathcal{G}}_{13}^0\tilde{\mathcal{G}}_{24}^0+\tilde{\mathcal{G}}_{12}^0\tilde{\mathcal{G}}_{14}^0\tilde{\mathcal{G}}_{23}^0-(\tilde{\mathcal{G}}_{12}^0)^2\tilde{\mathcal{G}}_{34}^0
\end{align}

\section{Disorder averaged coefficients for nodal voltages}\label{dis_coefficient_formulae}

We have the following integrals for the disorder averaged coefficients used to compute the nodal voltages---for one and two disordered bonds.  
\begin{subequations}
\begin{align}
\mathcal{I}_{11}^{[1]}&\equiv \int c_{11}^{[1]}\zeta_1  f(\zeta_1)\:  \d\zeta_1\\
\mathcal{I}_{11}^{[2]}&\equiv \iint c_{11}^{[2]}\zeta_1  f(\zeta_1) f(\zeta_2)\:  \d\zeta_1\: \d\zeta_2\\
\mathcal{I}_{22}^{[2]}&\equiv \iint c_{22}^{[2]}\zeta_2  f(\zeta_1) f(\zeta_2)\:  \d\zeta_1\: \d\zeta_2\\
\mathcal{I}_{12}^{[2]}&\equiv \iint c_{12}^{[2]}\sqrt{\zeta_1\zeta_2} f(\zeta_1) f(\zeta_2)\:  \d\zeta_1\: \d\zeta_2
\end{align}
\end{subequations}
where each of the integrals are integrated over the limits $0$ to $1-e^{-a}$. We also have $\mathcal{I}_{21}^{[2]}=\mathcal{I}_{12}^{[2]}$ by symmetry. The disorder averaged coefficients for three bonds can also be written similarly and are not provided here explicitly in view of conciseness.  

Finally, we also provide the four-point disorder average coefficients useful for calculating the nodal fluctuations---again, for one and two disordered bonds. 
\begin{subequations}
\begin{align}
\mathcal{J}_{1111}^{[1]}&\equiv \int \left(c_{11}^{[1]}\right)^2\zeta_1^2 f(\zeta_1)\:  \d\zeta_1\\
\mathcal{J}_{1111}^{[2]}&\equiv\!\iint\!\!\left(c_{11}^{[2]}\right)^2\zeta_1^2  f(\zeta_1) f(\zeta_2)\:  \d\zeta_1\: \d\zeta_2\\
\mathcal{J}_{2222}^{[2]}&\equiv\!\! \iint\! \left(c_{22}^{[2]}\right)^2\zeta_2^2 f(\zeta_1) f(\zeta_2)\:  \d\zeta_1\: \d\zeta_2\\
\mathcal{J}_{1212}^{[2]}&\equiv\!\! \iint\! \left(c_{12}^{[2]}\right)^2 \zeta_1\zeta_2  f(\zeta_1) f(\zeta_2)\:  \d\zeta_1\: \d\zeta_2\\
\mathcal{J}_{1122}^{[2]}&\equiv\!\! \iint\! c_{11}^{[2]}c_{22}^{[2]} \zeta_1\zeta_2  f(\zeta_1) f(\zeta_2)\:  \d\zeta_1\: \d\zeta_2\\
\mathcal{J}_{1112}^{[2]}&\equiv\!\! \iint\! c_{11}^{[2]}c_{12}^{[2]} \sqrt{\zeta_1^3\zeta_2}  f(\zeta_1) f(\zeta_2)\:  \d\zeta_1\: \d\zeta_2\\
\mathcal{J}_{2212}^{[2]}&\equiv\!\! \iint\! c_{12}^{[2]}c_{22}^{[2]} \sqrt{\zeta_1\zeta_2^3}   f(\zeta_1) f(\zeta_2)\:  \d\zeta_1\: \d\zeta_2
\end{align}
\end{subequations}
where the integrals are again over the limits $0$ to $1-e^{-a}$. The rest of the disorder averaged coefficients in $J_{\alpha\beta\gamma\delta}^{[n]}$ $(n=1,2)$ are all equal to the entries in the above set by symmetry. The coefficients for higher $n$ can be constructed similarly. 

\bibliography{refs}

\end{document}